\documentclass[
 reprint,
preprint,
 amsmath,amssymb,
]{elsarticle}

\usepackage{amsthm}

\usepackage[export]{adjustbox}
\usepackage{xcolor}
\usepackage[T1]{fontenc}
\definecolor{C0}{HTML}{1F77B4}
\definecolor{C1}{HTML}{FF7F0E}
\definecolor{C2}{HTML}{2ca02c}
\definecolor{C3}{HTML}{d62728}
\definecolor{C4}{HTML}{9467bd}
\definecolor{C5}{HTML}{8c564b}

\def\usetodonotes{} %% Comment this line to hide todos
\ifdefined\usetodonotes
\usepackage{todonotes}
\usepackage{changes}

\fi

\usepackage{upgreek}
\usepackage{tikz}
\usepackage[normalem]{ulem}
\usepackage{tcolorbox}
\usepackage[scr=esstix,cal=boondox]{mathalfa}

\usepackage{graphicx}% Include figure files
\usepackage{dcolumn}% Align table columns on decimal point
\usepackage{bm}
\usepackage[%Uncomment any one of the following lines to test 
margin=1in,
]{geometry}

\usepackage{siunitx}
\usepackage{siunitx}
\usepackage[skip=0pt]{caption}
\usepackage{subfig}
\usepackage{float}
\usepackage{amsmath}
\usepackage{amsmath}
\usepackage{amsfonts}
\newcommand{\angstrom}{\textup{\AA}}

\usepackage{comment}
\usepackage[english]{babel}
\usepackage{natbib}
\setcitestyle{square, numbers, comma, sort&compress, super}
\usepackage{algpseudocode}
\usepackage{algorithm}
\usepackage{latexsym}

\usepackage[colorlinks=true]{hyperref}
\hypersetup{ %Sets up hyperref
    colorlinks=true,       % false: boxed links; true: colored links
    linkcolor=red,          % color of internal links
    citecolor=blue,        % color of links to bibliography
    filecolor=magenta,      % color of file links
    urlcolor=cyan           % color of external links
}
\usepackage{cleveref}

\newcommand{\sref}[1]{Section~\ref{#1}}

\newcommand{\fref}[1]{Fig.~\ref{#1}}

\newcommand{\ms}{ms$^{-1}$ }
\newcommand{\emean}{$\epsilon_{mean}$ }

\begin{document}

%\preprint{APS/123-QED}

%\title{Interlayer Dislocations in Large Twist Bilayer Graphene and 2D Heterostructures
\title{Enhancement of plastic deformation in ultrasound-assisted cold spray of tungsten: a molecular dynamics study
}% Force line breaks with \\

\author[1]{Md Tusher Ahmed\corref{cor1}}
\ead{mdtusher.ahmed@utrgv.edu}
\cortext[cor1]{Corresponding author}
\author[1,2]{Farid Ahmed}
\author[1,2]{Jianzhi Li}%
\affiliation[1]{%
 Institute for Advanced Manufacturing, The University of Texas Rio Grande Valley, Edinburg, TX, USA
}%
%\affiliation[2]{%
%Department of Mechanical Engineering,
%Bangladesh University of Engineering and Technology, Dhaka, Bangladesh
%}%
\affiliation[2]{%
Department of Manufacturing and Industrial Engineering,
The University of Texas Rio Grande Valley, Edinburg, TX, USA
}%

%\affiliation{
%\affiliation{
%\affiliation{%


\begin{abstract}
Tungsten ($W$) is widely valued for its exceptional thermal stability, mechanical strength, and corrosion resistance, making it an ideal candidate for high-performance military and aerospace applications. However, its high melting point and limited room-temperature plasticity pose significant challenges for processing $W$ using additive manufacturing (AM). Cold spray (CS), a solid-state AM process that relies on high-velocity particle impact and plastic deformation, offers a promising route for additive manufacturing of $W$, yet conventional CS fails to induce sufficient plastic deformation for effective bonding. In this study, we employ atomistic simulations to investigate the effect of ultrasonic perturbation in enhancing plastic deformation during CS of $W$, with a focus on acoustoplasticity-driven deformation mechanism. We show that ultrasonic perturbation leads to pronounced acoustic softening and promotes transient temperature elevation at the particle-substrate and particle-particle interfaces, thereby enhancing plastic deformation compared to non-ultrasound-assisted CS. Additionally, our results show that the coupled effects of acoustic softening and enhanced transient thermal activation lead to substantial improvements in interfacial bonding across a wide range of impact velocities, particle sizes, and ultrasonic parameters. Finally, we analyze the feasibility of ultrasound-assisted CS for manufacturing heterogeneous interfaces consisting of an equimolar Vanadium ($V$)-Tungsten ($W$) coating on a $W$ substrate. Simulations reveal distinct mechanical behavior and dislocation densities compared to the homogeneous $W$ on $W$ CS configurations. Overall, this work highlights the potential of ultrasound-assisted cold spray as an effective strategy for manufacturing uniform coatings and engineered alloys, thereby addressing critical limitations in the additive manufacturing of refractory metals.
\end{abstract}

\maketitle

\section{\label{sec:level1}Introduction}
Cold spray additive manufacturing (CSAM) has revolutionized on-site surface repair through rapid, solid-state deposition of metal coatings \citep{widener2016application,yin2018cold,kanishka2023revolutionizing}. In cold spray (CS), metallic particles are accelerated by gas jets, which, upon striking the metal surface, cause plastic deformation, leading to the formation of a stable coating. The CS process overcomes the limitations of melting by operating at lower processing temperatures than traditional thermal-spray AM processes. In the absence of explicit melting, the cold-sprayed configurations exhibit lower thermal gradients and, consequently, lower residual stresses, as well as a lower tendency to form pores within the coating layers \citep{bagherifard2018cold}. CSAM is particularly popular for defense applications due to its ability to rapidly deploy in field-based defense scenarios, eliminating the need for controlled environments (e.g., inert gas chambers), laser alignment, and bulky equipment \citep{karthikeyan2004cold,widener2016application,sun2022current}. CS has been experimentally utilized for softer FCC materials, such as aluminum, nickel, and copper \cite{jasthi2023microstructure,chen2023ductile}, where these materials, in the form of particles, can undergo excessive plastic deformation, form dislocations, and grain boundaries due to impact with the product surface. However, quantification of plastic deformation, bond-formation parameters, and defect dynamics in harder BCC materials, such as tungsten ($W$), remains unexplored.    

Tungsten, $W$, is one of the hardest materials with a wide range of applications, spanning energy applications to defense manufacturing \citep{davis1998assessment,ueda2014research,arora2004tungsten,shedd2017tungsten}. Being a BCC material with inherently higher hardness, $W$ exhibits a brittle nature and sustains lower plastic deformation, forming higher porosity when subjected to high temperatures or external loads \citep{kafle2025review}. Moreover, spraying such heavy particles as $W$ requires a higher nozzle pressure ($>50$ bar) to typically create a velocity of $>1200$ \ms \citep{ALONSO2023103479}, making CSAM of $W$ economically unfeasible. Additionally, medium-pressure CSAM of $W$ can lead to larger pores due to its brittle nature, thereby affecting the stability of the final coating. Thus, despite having a wide range of applications, the manufacturing of $W$ coating using CS remains unexplored. Besides, even in high-pressure systems of tungsten, larger self-diffusion can not be realized \citep{petrovskiy2020analysis}, which inhibits the possibility of alloy formation using the CS process for $W$ \footnote{While some authors have experimentally shown the formation of alloys using CS \citep{chen2021cold,wu2022optimization}, there is no existing research work that highlights the mechanism of alloy formation of heavy metals such as $W$ to our knowledge.}. Here, we present a novel technique to enhance the plastic deformation of $W$ particles on a $W$ substrate with the application of ultrasonic perturbation using atomic-scale simulations. In addition, we demonstrate, at the atomic level, how ultrasound can induce transient thermal activation, leading to the intermixing of dissimilar atoms and thereby expanding the scope of alloy-based coating formation using the CSAM.  While the application of ultrasound has been explored for removing oxide \citep{long2019impacts} in various AM processes, ultrasonic perturbation's effect on deformation behavior and defect dynamics has not been investigated in the CS process yet \footnote{While effect of ultrasound has been explored for FCC metals in laser assisted additive manufacturing processes, \cite{zhang2024advances,todaro2021grain,li2024ultrasonic}, effect of ultrasound in CSAM has not been unexplored yet.}.

Computational modeling of the CS plays a crucial role in gaining insight into particle-substrate interactions during the CS process. Several continuum-scale research works have captured interesting physical phenomena, such as mechanical interlocking, adiabatic shear instability, and interface diffusion, during the CS process \citep{tianyu2022experimental,zorawski2021experimental,deng2023simulation}. However, bond formation and microstructural evolution during the CS process occur at a much smaller timescale than the accessible timescale of continuum-scale methods \citep{zhang2025multi}. These microstructural changes can be actively captured by the molecular dynamics (MD) simulation techniques \citep{joshi2018molecular,gao2023tamping}. While existing studies on MD simulations have focused on softer materials, such as copper \citep{feng2024atomistic,gao2023tamping} and nickel\citep{wu2025exploring}, as well as medium-hard materials, such as titanium \citep{dai2025molecular}, atomistic details about interfacial bond formation, deformation mechanisms, and defect dynamics in $W$, one of the hardest materials, remain unexplored.    

The paper is organized as follows. \sref{sec:model_des} presents the methodology for the MD simulation of the CS process. \sref{sec:visco_800} shows how microstructural evolution occurs as a function of time at a specific velocity and particle size in the presence and absence of ultrasonic perturbation. To ensure the generality of the findings observed in \sref{sec:visco_800}, we investigate the effect of different CS process parameters on the deformation behavior, diffusive nature, and morphological evolution of the particles during the CS process in \sref{sec:param}. In \sref{sec:alloy_form}, we analyze bonding, deformation, and defect dynamics of $V$-$W$ alloy heterogeneous coating on a $W$ substrate, manufactured using ultrasound-assisted CS process. We summarize and conclude in \sref{sec:summary}. 

\section{Model}
\label{sec:model_des}
\begin{figure}
    \centering
    \includegraphics[width=0.7\linewidth]{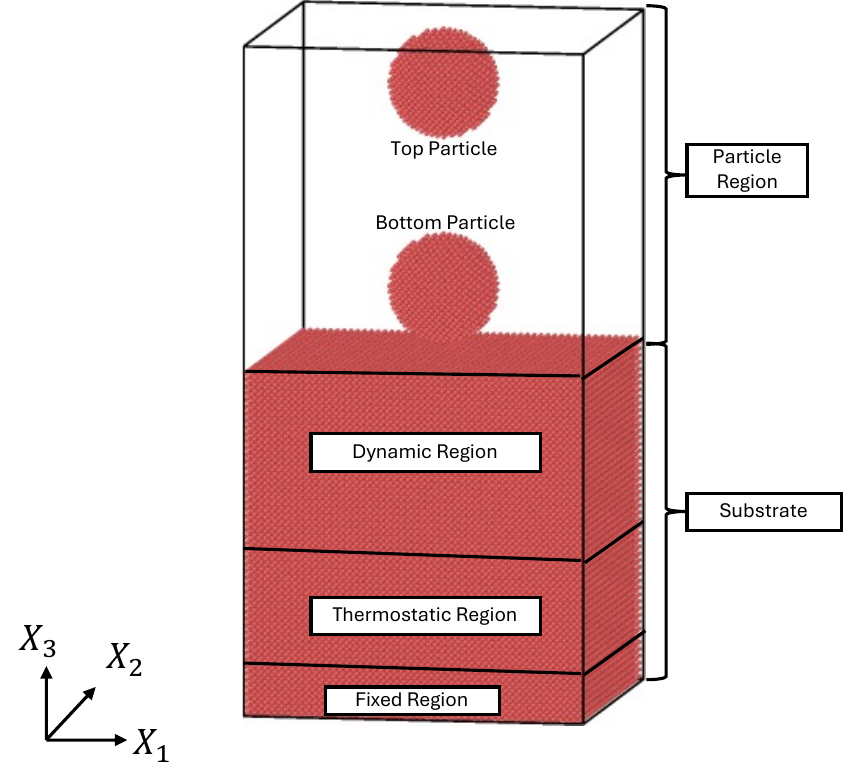}
    \caption{Schematic diagram of the cold spray simulation}
    \label{fig:schematics}
\end{figure}

We use atomistic simulations to understand the underlying physics of the non-ultrasound and ultrasound-assisted CS process of $W$. Simulations are conducted using the Large-scale Atomic/Molecular Massively Parallel Simulator (LAMMPS) \citep{LAMMPS}. The schematic of the simulation setup is shown in \fref{fig:schematics}. A simulation domain has been designed with periodic boundary conditions in the $X_1$-$X_2$ plane. At the same time, a shrink-wrapped boundary condition has been applied along the $X_3$ direction to allow free motion of the particles in this direction. The cubic substrate dimension is kept to be $158.25\si{\angstrom}$ along the global coordinate axis where $X_1,X_2$ and $X_3$ correspond to $[100],[010]$ and $[001]$ directions respectively. Interaction between $W$ atoms is modeled using the Embedded Atom Method (EAM) potential developed by \citet{chen2020interatomic}. One of the benefits of this potential is its ability to model both $W$ and $V$-$W$ alloys accurately. The simulation domain is divided into two regions: the Substrate Region and the Particle Region. While the substrate mimics the part to be repaired, particles mimic the incoming CS coating materials that will be deposited. To allow the top of the substrate to be governed solely by the dynamics of the particle-substrate interaction, we divided the substrate into three segments. The bottom-most segment of the substrate, with a height of $31.65\si{\angstrom}$, is modeled as the fixed region, where all degrees of freedom are set to zero for the non-ultrasound-assisted case. The next $47.48\si{\angstrom}$ segment is modeled as the thermostatic region where the NVT canonical ensemble is maintained at the temperature of $300$K using the Nos\'e-Hoover thermostat \citep{parrinello1981polymorphic}. The topmost region of the substrate is defined as the dynamic region. The NVE microcanonical ensemble is maintained to model dynamics due to particle-substrate interactions. In comparison to the non-ultrasound-assisted case, the fixed region in the ultrasound-assisted configuration is subjected to a harmonic displacement of the following form,
\begin{equation}
    X\left(t\right)=X_0+A\sin(2\pi f \Delta t)
\end{equation} where, $A=$ the amplitude of the perturbation and, $f=$ the frequency of the perturbation. Here, the perturbation is applied along the out-of-plane $X_3$ direction to simulate a vertical ultrasound effect. However, the in-plane degrees of freedom of this fixed region are kept to zero. This particular approach has been adapted by researchers specifically in the context of ultrasonic friction stir welding of dissimilar joints \cite{yang2019molecular,yang2024effect}. The particle region consists of two spherical particles with a diameter of $50.64\si{\angstrom}$ to mimic the CS process with the tamping effect \cite{li2003deposition}. The purpose of launching two particles along the same trajectory is to maximize the plastic deformation induced by the tamping effect, an approach implemented by \citet{gao2023tamping,zhang2025multi} in their studies, based on the observation of linearity in the flattening ratio with the number of tamping layers. Centers of the particles are placed $37.98\si{\angstrom}$ and $164.58\si{\angstrom}$ above the substrate, respectively. Before running the dynamics calculation, we relax the configuration using the conjugate gradient minimization algorithm. Following this, a $0.1$ ns-long dynamic equilibrium simulation has been conducted without the initiation of impact. At time $t=0.1$ns, the dynamics simulation begins with the impact of the bottom particle at an impact velocity of $800$ \ms, which is followed by the impact of the top particle at $t=0.3$ ns. The time delay of $0.2$ ns between successive impacts allows the deformation measures to reach a steady state following the first impact before the collision with the second particle. Moreover, the particle spacing in this study has been carefully selected to ensure that the bottom particle's impact on the substrate does not affect the top particle until the top particle strikes at time $t=0.3$ ns. The particles are subjected to the NVE ensemble during the entire simulation. 

\section{Ultrasound Assisted Acoustoplasticity in Tungsten}
\label{sec:visco_800}
The Cold Spray process involves the deposition of nano- to micro-sized particles on substrate materials through high-velocity impact. The particles' kinetic energy is converted into localized plastic deformation. Moreover, intermixing and metallurgical bonding between the particle and substrate are also observed at the atomic scale  \cite{wu2025atomic}. The plastic deformation is primarily governed by dislocation generation, propagation, and annihilation at the particle and substrate region \cite{joshi2018molecular}. Additionally, during the tamping process, in which multiple particles are sequentially impacted, overlapping plastic zones lead to cumulative dislocation networks and interatomic bond formation across particle–particle and particle–substrate interfaces \cite{zhang2025multi}.
However, as one of the hardest materials, $W$ is difficult to process using CSAM due to its low plasticity and brittle behavior under loading and impact. In this section, we present how ultrasonic perturbation applied to the substrate can amplify the plastic deformation achievable in a CS process.

\begin{figure}[t!]
    \centering
    \subfloat[]
        {
\includegraphics[height=0.51\textwidth]{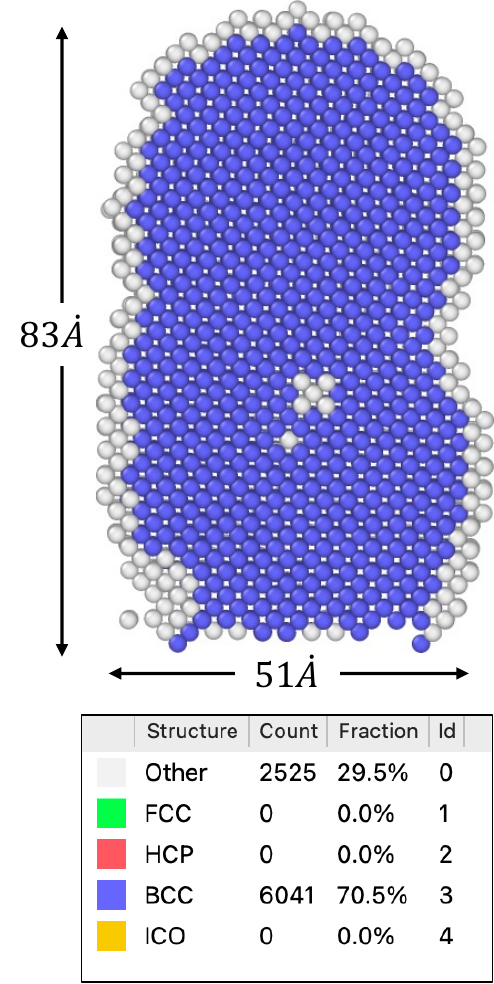}
    \label{fig:800_no_ultra_cal}
    }
    \subfloat[]
    {
\includegraphics[height=0.48\columnwidth]{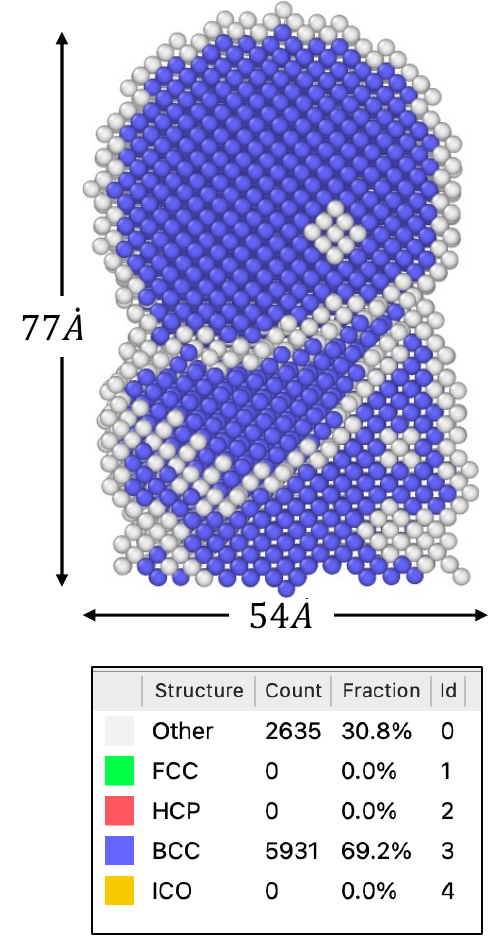}
        \label{fig:800_ultra_cal}
    }
    \caption{Variation of atomic configurations of $W$ particles and substrate after $t=0.5$ ns following an impact at a velocity, $v=800$ \ms \protect\subref{fig:800_no_ultra_cal} without ultrasonic perturbation and, \protect\subref{fig:800_ultra_cal} subjected to an ultrasonic perturbation with the amplitude of $A=3.165\si{\angstrom}$ and frequency, $f=10$ GHz. The number of different lattice structures and dislocations corresponding to each atomic configuration is shown below the figures. The complete dynamic evolution of CS particles following impact can be observed in the supplementary videos.}
    \label{fig:800_begin}
\end{figure}

\begin{figure}[t!]
    \centering
    \subfloat[]
    {
\includegraphics[height=0.4\textwidth]{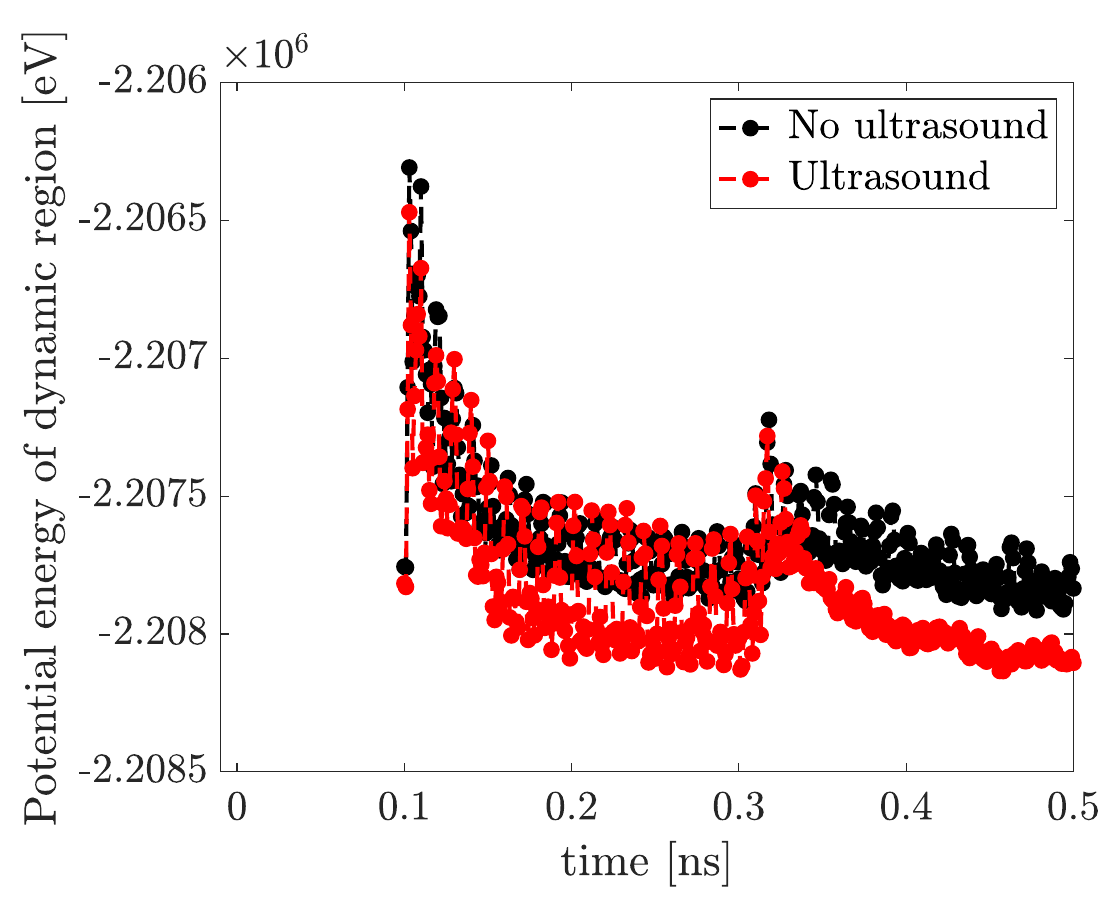}
    \label{fig:energy_track}
    }
    \subfloat[]
    {
\includegraphics[height=0.4\columnwidth]{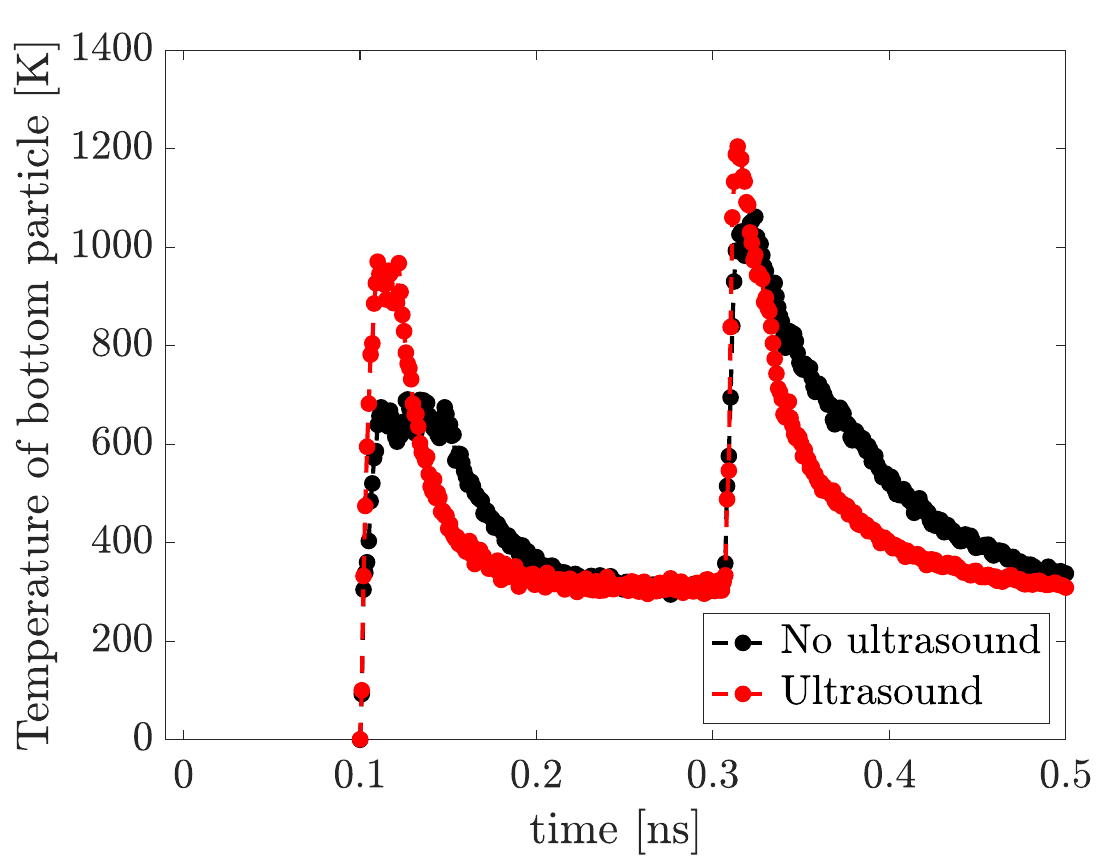}
        \label{fig:temp_track}
    }
    %\\  
    %\subfloat[]
    %{
%\includegraphics[height=0.4\columnwidth]{Temp_track_dynamic.pdf}
%        \label{fig:temp_track_dyn}
%    }
    \caption{Variation of \protect\subref{fig:energy_track} substrate potential energy and \protect\subref{fig:temp_track} temperature of bottom particle %\protect\subref{fig:temp_track_dyn} temperature of dynamic region
    as a function of time at non-ultrasound and ultrasound-assisted cases.}
    \label{fig:en_temp_track}
\end{figure}

To investigate the self-diffusion in $W$, we first present the atomic configurations (plotted using OVITO \citep{ovito}) of the particle region at time $t=0.5$ ns in the presence and absence of ultrasonic perturbation. Left configuration in \fref{fig:800_ultra_cal} illustrates the microstructure of the configuration in the absence of ultrasonic perturbation, while the right one illustrates the configuration in the presence of ultrasonic perturbation. We note from \fref{fig:800_no_ultra_cal} that a vertical compression of $\approx18\si{\angstrom}$ occurs along $X_3$ direction for the bottom particle after impact in the absence of ultrasonic perturbation. However, \fref{fig:800_ultra_cal} shows a vertical compression of $\approx23\si{\angstrom}$ when the fixed region is subjected to ultrasonic perturbation of an amplitude of $A=3.165\si{\angstrom}$ and frequency, $f=10$ GHz \footnote{Unless otherwise stated, ultrasonic perturbation refers to this periodic displacement perturbation with an amplitude of $A=3.165\si{\angstrom}$ and frequency, $f=10$ GHz}. The ultrasonic perturbation enhances the tendency of grain refinement and recrystallization \cite{li2025effect} as observed in \fref{fig:800_ultra_cal} through the formation of multiple grain boundaries. Such grain refinement is assisted from the increment in temperature of the bottom particle as shown in \fref{fig:temp_track} in ultrasound-assisted case which is 400K higher than the corresponding non-ultrasound case at time $t\approx0.1$ ns. While the initial impact increases the temperatures significantly, the system soon reaches to a thermal equilibrium due to the interaction with the thermostat. Moreover, as observed in \fref{fig:energy_track}, the continuous perturbation of the fixed region does not result in any artificial energy surge. Additionally, ultrasonic perturbation leads to grain refinement and defect rearrangements, resulting in a lower overall potential energy than in the non-ultrasound-assisted case. Furthermore, the variation in BCC atom counts shown in \fref{fig:800_begin} for the non-ultrasound and ultrasound-assisted cases also illustrates enhanced atomic rearrangements during the ultrasound-assisted CS process.

\begin{figure}[t!]
    \centering
    \subfloat[MSD-No Ultrasound]
    {
\includegraphics[width=0.5\textwidth]{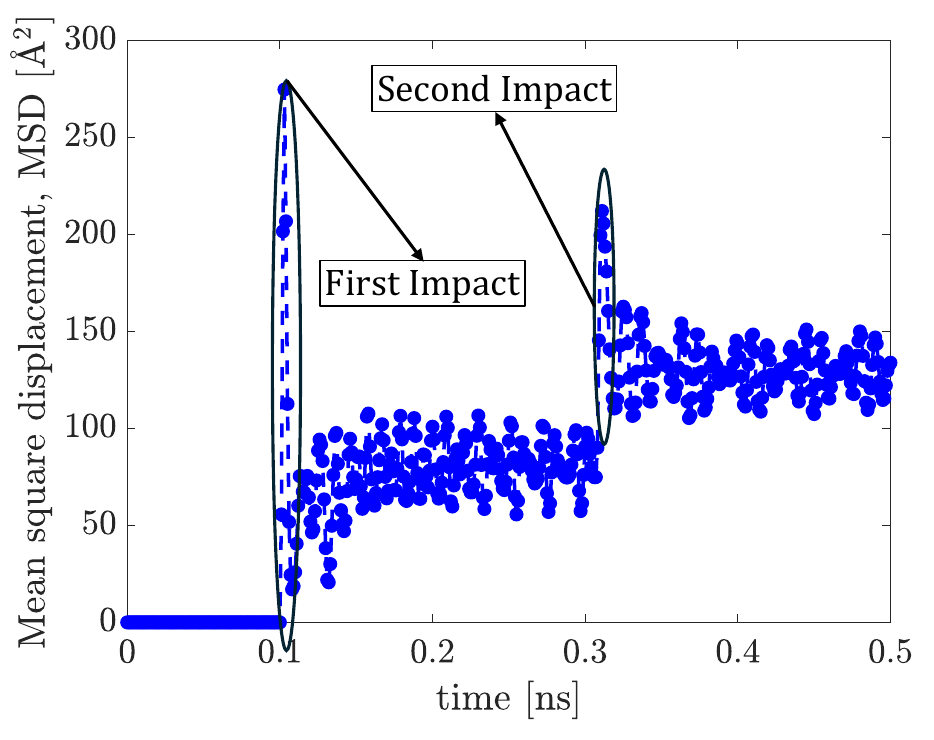}
    \label{fig:800_no_ultra_MSD}
    }
    \subfloat[MSD-Ultrasound]
    {
\includegraphics[width=0.5\columnwidth]{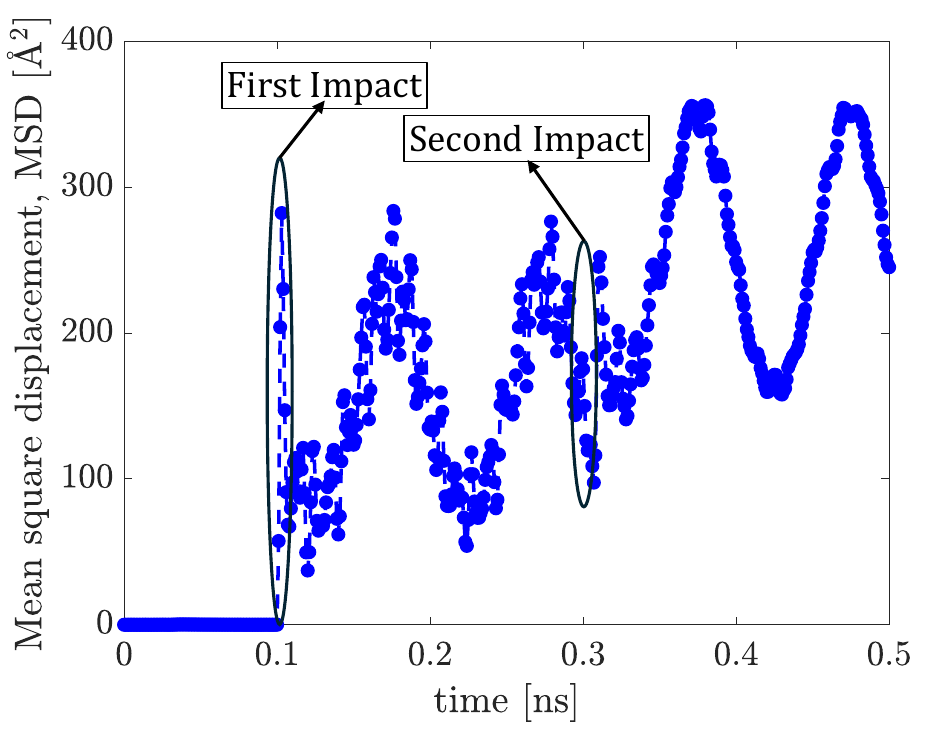}
        \label{fig:800_ultra_MSD}
    }
    \caption{Variation of mean square displacement of the bottom $W$ particle as a function of time in the presence and absence of ultrasonic perturbation with the amplitude of $A=3.165\si{\angstrom}$ and frequency, $f=10$ GHz at an impact velocity of $v=800$ \ms.}
    \label{fig:800_param}
\end{figure}

\begin{figure}[t!]
    \centering
    \subfloat[No Ultrasound]
    {
\includegraphics[width=0.5\textwidth]{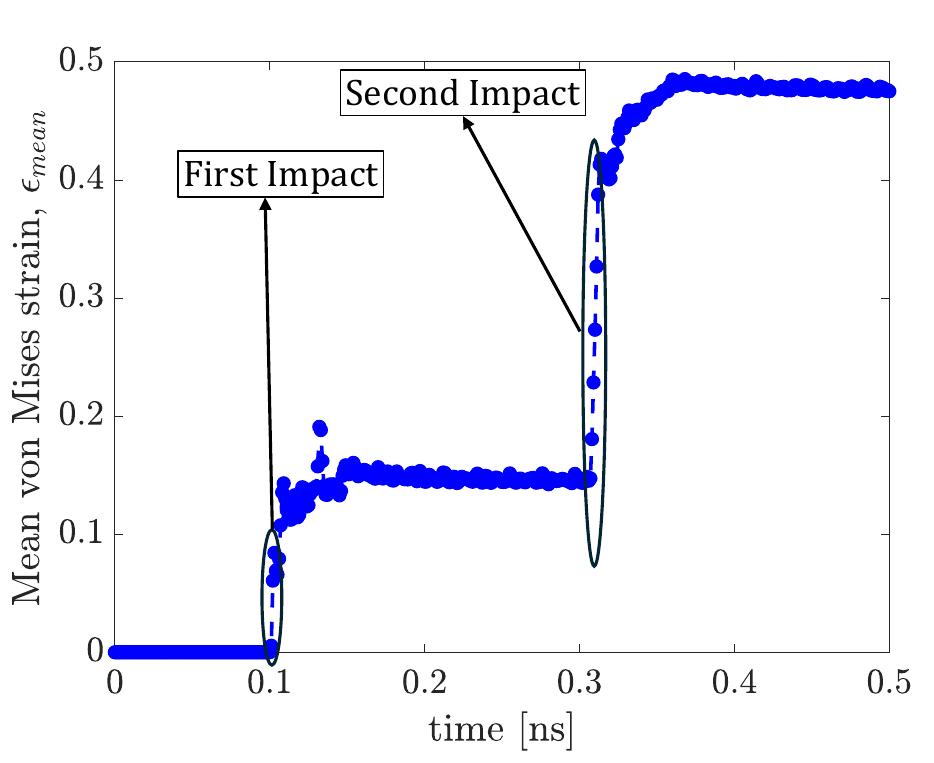}
    \label{fig:800_no_ultra_mean_str}
    }
    \subfloat[Ultrasound]
    {
\includegraphics[width=0.5\columnwidth]{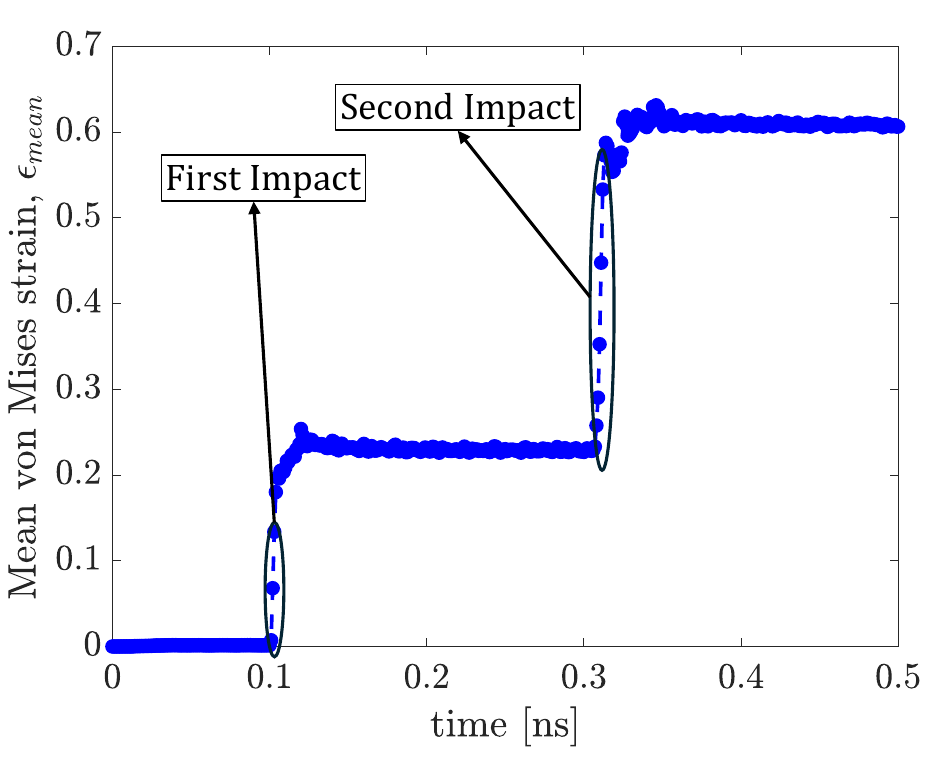}
        \label{fig:800_ultra_mean_str}
    }
    \caption{Variation of mean von Mises strain of the bottom $W$ particle as a function of time in the presence and absence of ultrasonic perturbation with the amplitude of $A=3.165\si{\angstrom}$ and frequency, $f=10$ GHz.}
    \label{fig:800_begin_mean}
\end{figure}

To quantify the self-diffusion within $W$ particles on $W$ substrate in the absence and presence of ultrasonic perturbation, we computed the mean square displacement (MSD) of the bottom particle at various times as the simulation progressed. \fref{fig:800_no_ultra_MSD} shows the evolution of MSD of the bottom particle as a function of time in the absence of ultrasound. It can be observed that the MSD exhibits high-frequency, large-amplitude oscillations immediately after the impact at time $t=0.1$ ns, followed by saturation within a short interval. A similar saturation behavior of the MSD is observed following the second impact at $t=0.3$ ns, albeit at a higher value. As the MSD saturates following impact, the diffusion is considered absent in the non-ultrasound case. In contrast, \fref{fig:800_ultra_MSD} displays an oscillatory evolution of the MSD over time, matching the frequency of the applied ultrasonic perturbation. However, no noticeable increase in the mean MSD is observed, except during impact events. This suggests that, as in the non-ultrasound-assisted case, diffusion remains absent in the ultrasound-assisted case at $800$ \ms. Therefore, the observed lateral compression is not a direct result of enhanced diffusion induced by the ultrasonic perturbation.
To further investigate the enhancement of lateral compression at the ultrasound-assisted configuration, we computed the atomic von Mises strain \footnote{Atomic strain is a local measure of distortion of an atom’s neighborhood relative to a reference lattice \cite{pyrz2006atomistic}, which in this case is the minimized configuration at time $t=0$. It is different from continuum macroscopic strain but correlates strongly with plastic deformation and dislocation activity, as shown by \citet{shimizu2007theory}. A high value of atomic strain indicates a high level of localized plastic deformation in a continuum body.} of the bottom particle atoms using the following expression.
\begin{equation}
    \epsilon=[E_{xy}^2+E_{yz}^2+E_{xz}^2+\frac{1}{6}\{(E_{xx}-E_{yy})^2+(E_{yy}-E_{zz})^2+(E_{zz}-E_{xx})^2\}]
    \label{eqn:str_von}
\end{equation}
where, $E_{ij}$ corresponds to the strain components of the 3D strain tensor.
\fref{fig:800_begin_mean} compares the deformation behavior of the bottom particle as a function of time in the absence and presence of ultrasonic perturbation \footnote{To evaluate the simulation size and timestep independence, simulations have been conducted for three different-sized simulation boxes and three different time steps. The findings are presented in Supporting Information S-1.}. Similar to the evolution of MSD in the non-ultrasound-assisted case, as shown in \fref{fig:800_no_ultra_mean_str}, the saturation of the mean von Mises strain, \emean, is evident following the impact events. Such saturation of \emean indicates the presence of a rate-independent plastic deformation mechanism in the non-ultrasound-assisted case. The maximum value of \emean is found to be $0.48$ only. However, \fref{fig:800_ultra_mean_str} shows an increase in the maximum value of \emean from $0.48$ to $0.60$ when subjected to ultrasonic perturbation. Moreover, a linear increase in \emean is observed for a short time interval after individual impact events. The modification in deformation behavior is a consequence of the combined effects of acoustoplasticity and the additional heating caused by the ultrasonic perturbation, as experimentally demonstrated by \citet{siu2011new} in aluminium. The ultrasonic excitation lowers the effective flow stress while raising the temperature of the collision interface, leading to grain refinement, atomic rearrangements, and additional plastic deformation. This behavior, known as acoustic softening, was first demonstrated by \citet{blaha1955tensile}.

\begin{figure}[t!]
    \centering
\includegraphics[width=0.5\columnwidth]{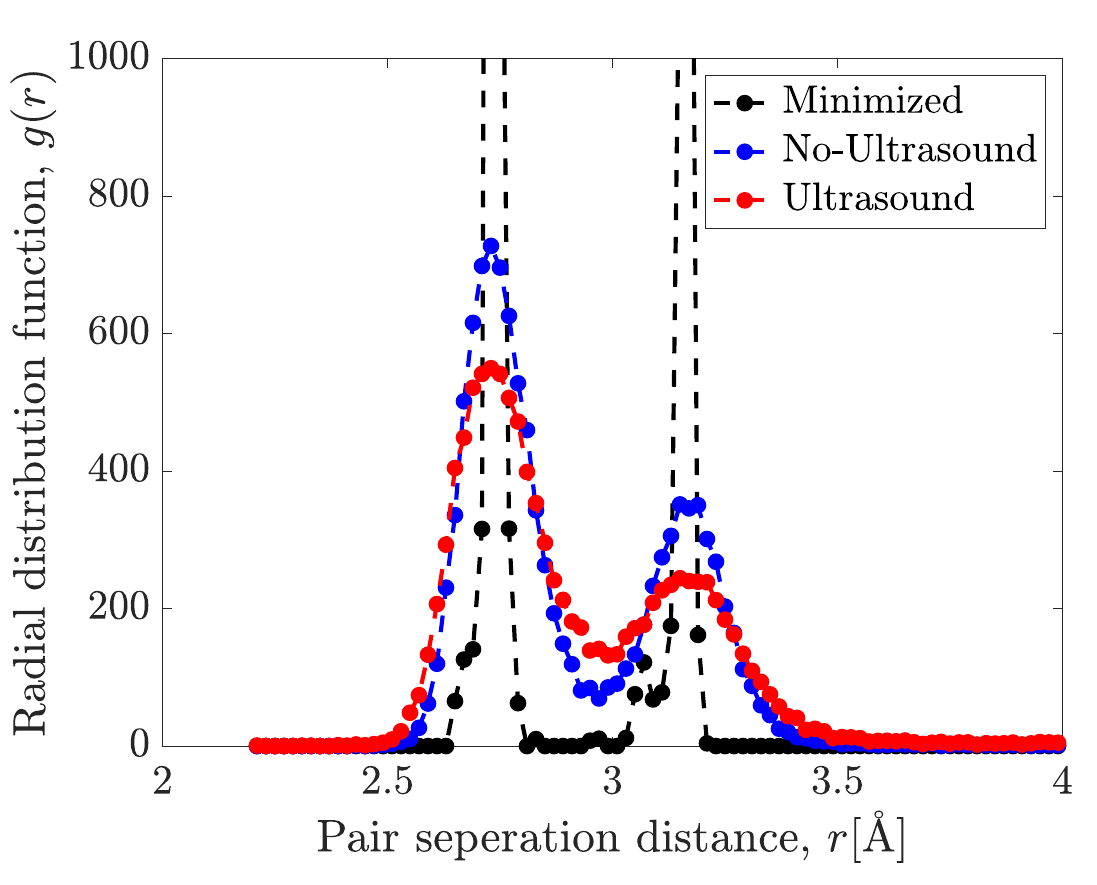}
    \caption{Variation of radial distribution functions of the bottom $W$ particle at time, $t=0.11$ ns in the presence and absence of ultrasonic perturbation with the amplitude of $A=3.165\si{\angstrom}$ and frequency, $f=10$ GHz.}
    \label{fig:800_rdf_calc}
\end{figure}

While MSD and \emean can provide insights into diffusion and deformation mechanisms, they cannot reveal the long-range order that defines the structure's physical state. To understand the crystallographic origin of this ultrasound-assisted acoustoplasticity in $W$, we computed the instantaneous radial distribution function (RDF) of the bottom $W$ particle for both non-ultrasound and ultrasound-assisted cases at time $t=0.11$ ns following the first impact using OVITO. RDF can be an effective tool in identifying the long-range order of a crystalline structure when subjected to added external perturbations. \fref{fig:800_rdf_calc} compares the ultrasound-assisted and non-ultrasound-assisted configurations' RDF distributions for the bottom particle at time $t=0.11$ ns with the minimized bottom particle RDF distribution. We note that the RDF overall corresponds to a crystalline BCC lattice structure with two peaks at $\frac{\sqrt{3}a}{2}$ and $a$ respectively, where $a$ is the lattice parameter of $W$. The difference in the width of the RDF peaks between the minimized and configuration at time $t=0.11$ ns is an alternate representation of the associated plastic deformation. Moreover, the peaks observed in ultrasound-assisted cases are slightly wider compared to those in non-assisted cases. This illustrates that ultrasonic perturbation induces acoustic softening \citep{mao2020investigating} in the $W$ particles, leading to the formation of a uniform coating on the substrate.  

Finally, to compare the quality of the deposited coating on the substrate between the non-ultrasound and ultrasound-assisted cases, we define a characteristic non-dimensional quantity, the maximum flattening ratio, $FR_{max}$. $FR_{max}$ can be expressed using the following equation,
\begin{equation}
    FR_{max}=1-\frac{T_{min}}{D_{init}}
\end{equation}
where $T_{min}$ and $D_{init}$ correspond to the minimum thickness of the bottom particle along $X_3$ direction following impact at time, $t=0.5$ ns, and the diameter of the bottom particle before impact at time, $t=0$, respectively. $FR_{max}$ quantifies the amount of deformation the bottom particle undergoes due to the impact with the substrate and top particles. \footnote{$FR_{max}$ only refers to the maximum compression the bottom particle undergoes due to impact with the top particle and substrate. However, this can overestimate the value of the flattening ratio.} At an impact velocity of $800$ \ms, $FR_{max}$ reaches a value of $0.26$, only showing a change of $\frac{1}{4}$th of the initial diameter without ultrasonic perturbation. However, the value increases to $0.40$ under ultrasonic perturbation. This demonstrates the enhanced dispersion of the coating materials under ultrasonic perturbation, resulting in a flatter, more uniform coating on the substrate surface. 

While this section presents an argument and supporting evidence behind the observation of acoustic softening-induced plasticity in ultrasound-assisted CS configuration at $800$ \ms for $50.64\si{\angstrom}$ particles under ultrasonic perturbation of an amplitude of $A=3.165\si{\angstrom}$ and frequency, $f=10$ GHz, we explore the generality of such a phenomenon at different impact velocities, particle dimensions, and frequencies in the next section.

\section{Effect of process parameters on ultrasound-assisted CSAM}    
\label{sec:param}
In the previous section, we demonstrated ultrasound-assisted acoustic softening in tungsten ($W$) at a specific impact velocity of $800$ \ms and particle size of $D=50.64\si{\angstrom}$. To assess the generality of this behavior, this section examines the influence of key process parameters on the post-impact deformation and diffusion characteristics of cold-sprayed particles. By systematically varying the impact velocity, particle size, and ultrasound frequency, we aim to gain a deeper understanding of how these parameters influence strain accumulation, atomic mobility, and the resulting microstructural evolution.
\subsection{Impact velocity}
\label{sec:impac_vel}
\begin{figure}[t!]
    \centering
    \subfloat[$400$ \ms-No Ultrasound]
    {
\includegraphics[width=0.28\textwidth]{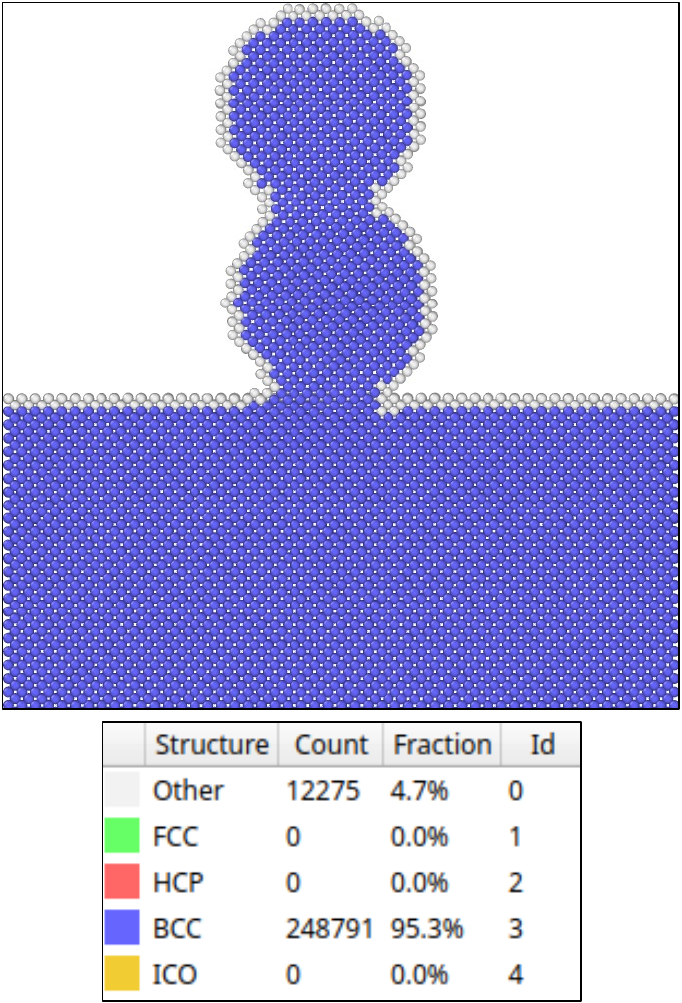}
    \label{fig:400_no_ultra_config}
    }
    \subfloat[$800$ \ms-No Ultrasound]
    {
\includegraphics[width=0.28\textwidth]{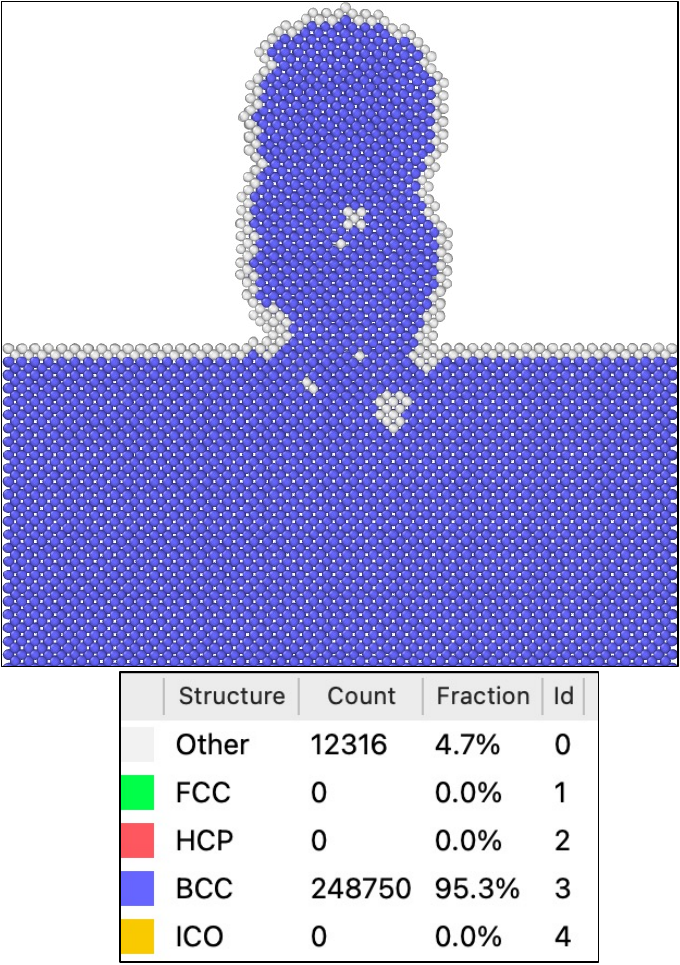}
    \label{fig:800_no_ultra_config}
    }
    \subfloat[$1200$ \ms-No Ultrasound]
    {
\includegraphics[width=0.28\textwidth]{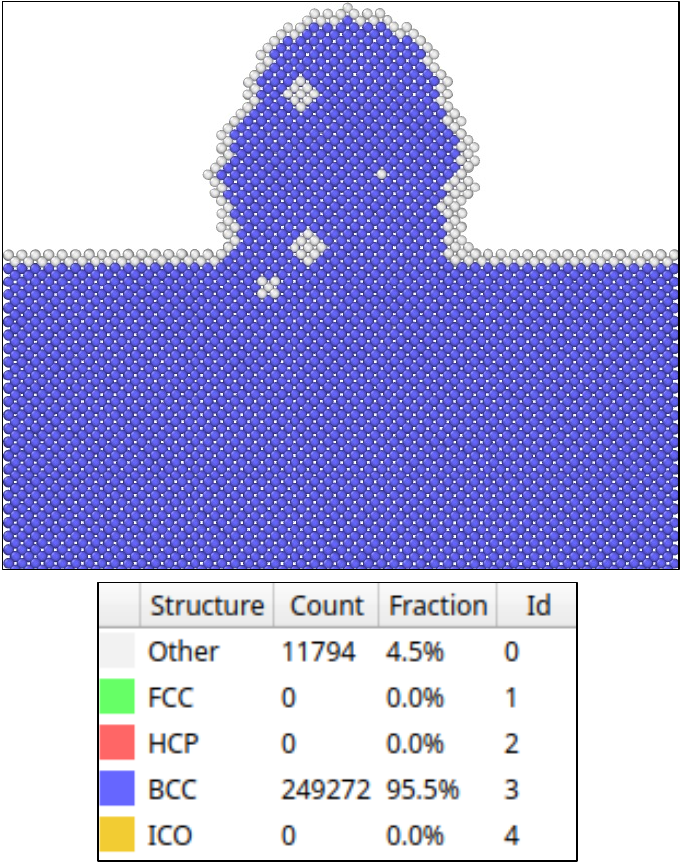}
    \label{fig:1200_no_ultra_config}
    }
    \\
    \subfloat[$400$ \ms-Ultrasound]
    {
\includegraphics[width=0.28\textwidth]{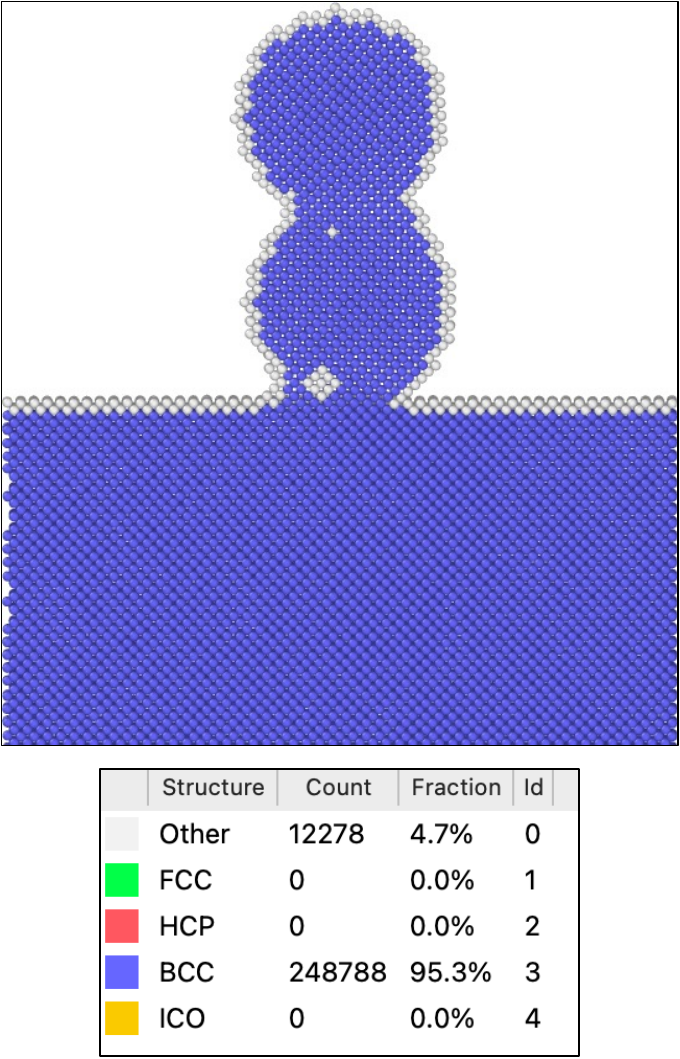}
    \label{fig:400_ultra_config}
    }
    \subfloat[$800$ \ms-Ultrasound]
    {
\includegraphics[width=0.28\textwidth]{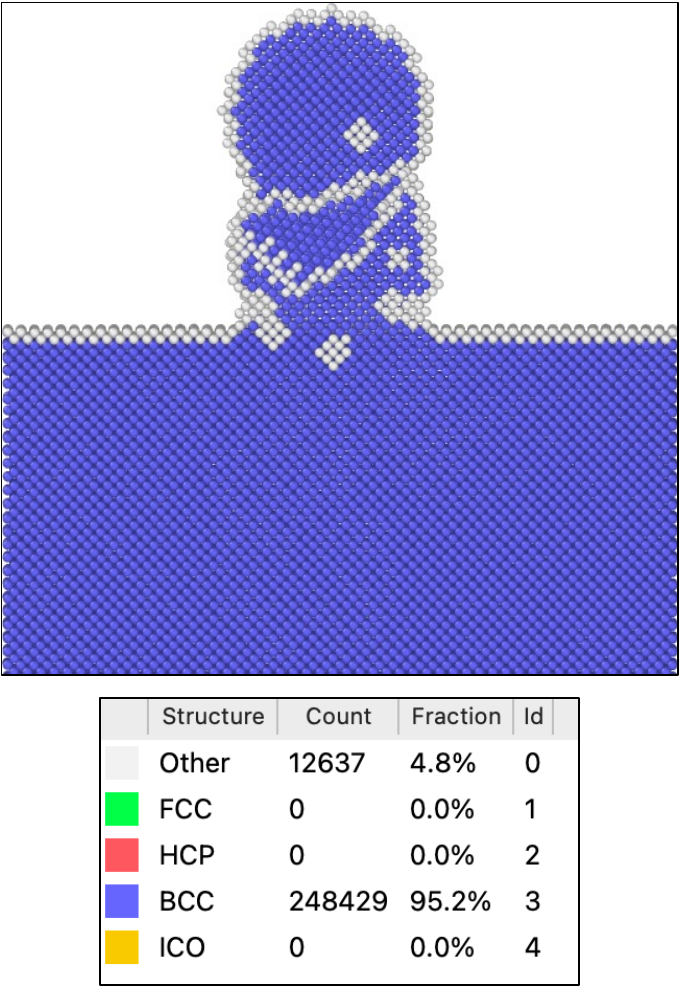}
    \label{fig:800_ultra_config}
    }
    \subfloat[$1200$ \ms-Ultrasound]
    {
\includegraphics[width=0.28\textwidth]{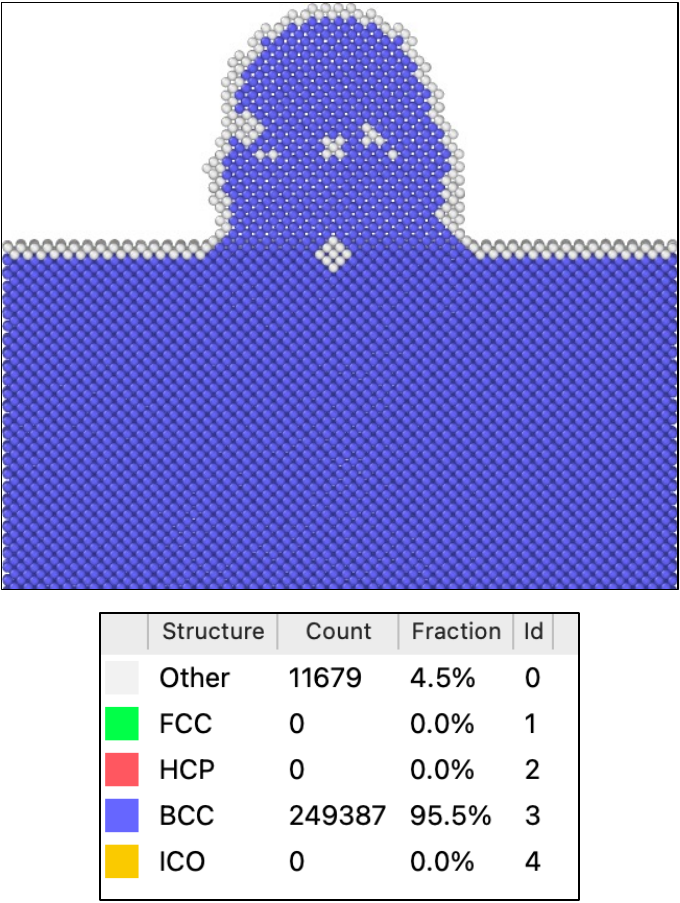}
    \label{fig:1200_ultra_config}
    }
    \caption{Variation of atomic configurations of $W$ particles and substrate after $t=0.5$ ns following impacts at different velocities in the absence [\protect\subref{fig:400_no_ultra_config},\protect\subref{fig:800_no_ultra_config} and\protect\subref{fig:1200_no_ultra_config}] and presence [\protect\subref{fig:400_ultra_config},\protect\subref{fig:800_ultra_config} and\protect\subref{fig:1200_ultra_config}] of ultrasonic perturbation with the amplitude of $A=3.165\si{\angstrom}$ and frequency, $f=10$ GHz. The number of different lattice structures corresponding to each atomic configuration is shown below the figures.}
    \label{fig:vel_variation_config}
\end{figure}

We first investigate the effect of impact velocity on the deformation and diffusion behavior of the bottom particle, both in the absence and in the presence of ultrasonic perturbation.  The velocity has been varied from $300$ to $1200$ \ms to explore a wide range of velocities accessible in cold-spray experiments \citep{indu2024quantitative}. To analyze the effect of impact velocities on the microstructure of the $W$ particles, we illustrate the atomic configurations within the simulation domain in \fref{fig:vel_variation_config}. \fref{fig:vel_variation_config} compares atomic configurations of the cold-sprayed system at time $t=0.5$ ns in the absence and presence of ultrasonic perturbation under three different impact velocities. In non-ultrasound cases illustrated by \fref{fig:400_no_ultra_config}, \fref{fig:800_no_ultra_config}, and \fref{fig:1200_no_ultra_config}, there is an increment in the percentage of BCC lattices as the impact velocity increases from $800$ \ms to $1200$ \ms. The percentage of the BCC lattice within the configuration increases by $0.2\%$ as the velocity increases from $800$ \ms to $1200$ \ms. Moreover, the height of the coating on the $W$ substrate in the $X_3$ direction decreases as the impact velocity increases. On the contrary, \fref{fig:400_ultra_config}, \fref{fig:800_ultra_config}, and \fref{fig:1200_ultra_config} illustrate the atomic configurations in the presence of ultrasonic perturbations. It can be observed that the crystallinity of the configurations decreases at impact velocities of $400$ and $800$ \ms, under ultrasonic perturbation. This behavior corresponds to the acoustic softening of the CS particles under ultrasonic perturbation. At such impact velocities, grain refinement and atomic rearrangement rate could be lower, reducing the amount of BCC lattice. However, at higher velocities, such as $1200$ \ms, the recrystallization rate could be high enough due to high transient temperature, leading to an increase in the BCC lattice fraction.  

\begin{figure}[t!]
    \centering
    \subfloat[$\epsilon_{mean}$-No Ultrasound]
    {
\includegraphics[width=0.5\textwidth]{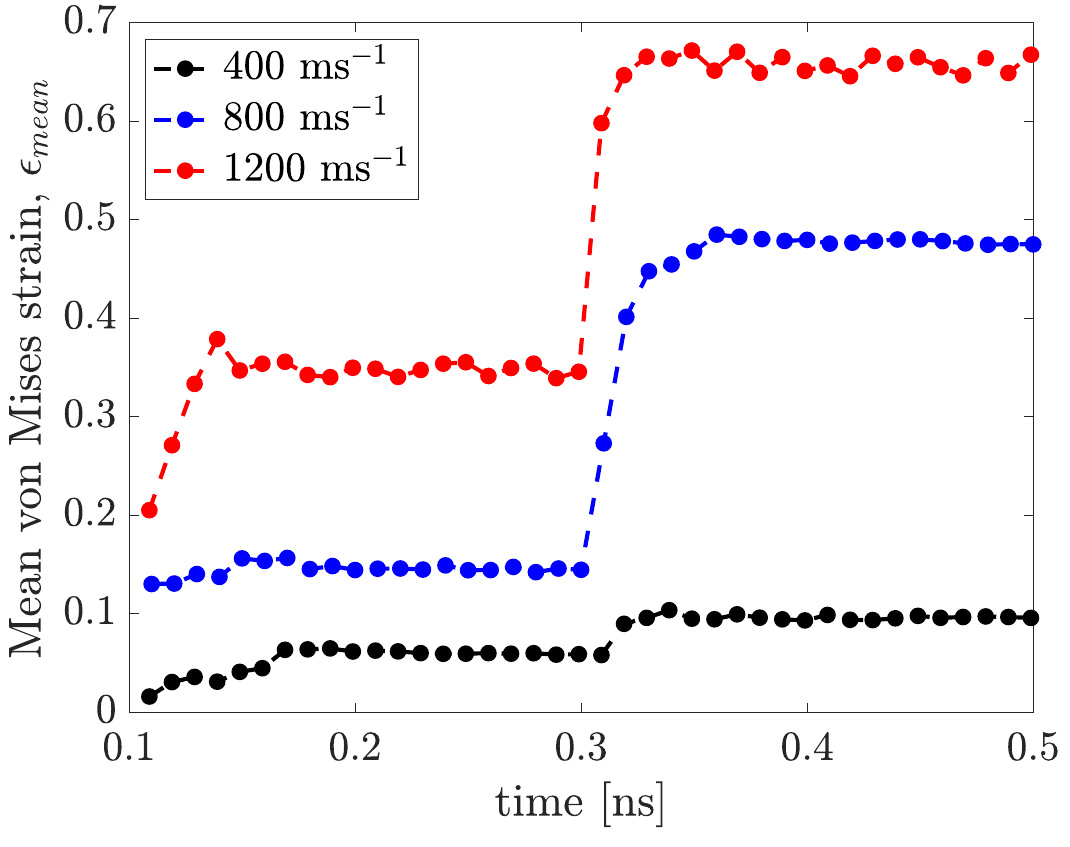}
    \label{fig:mean_str_nu_vel_var}
    }
    \subfloat[$\epsilon_{mean}$-Ultrasound]
    {
\includegraphics[width=0.52\textwidth]{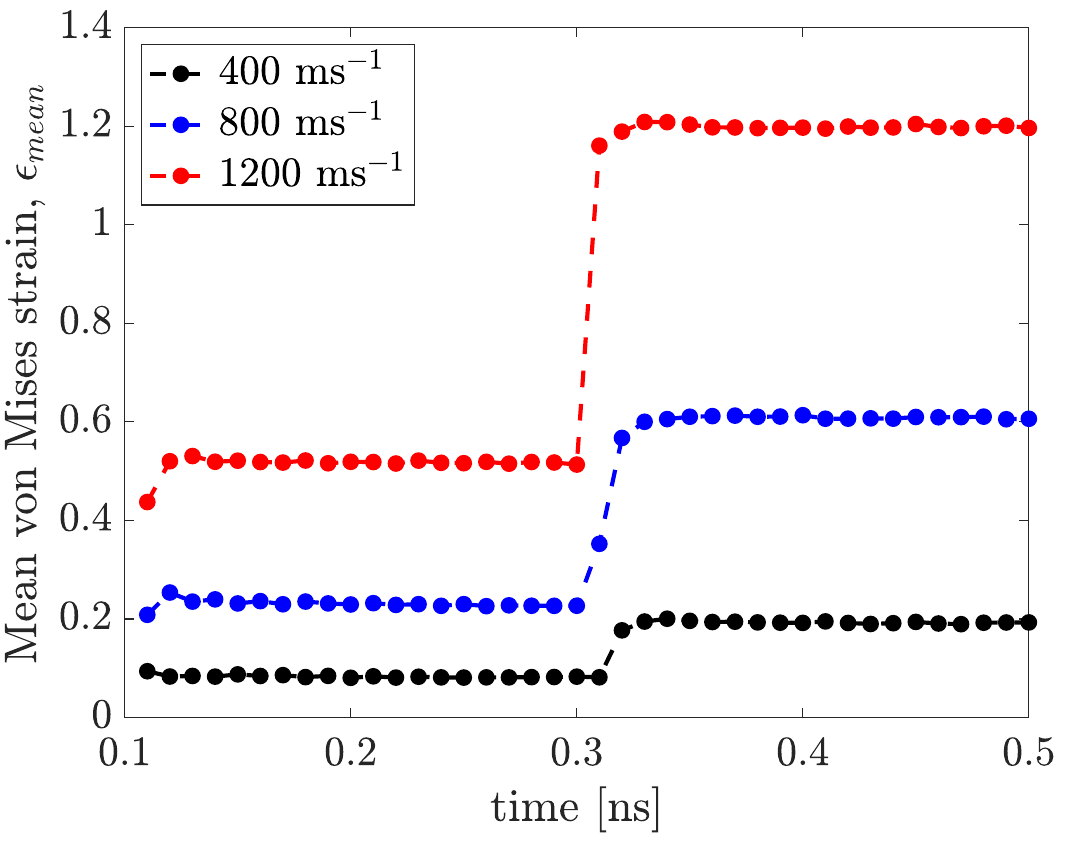}
    \label{fig:mean_str_u_vel_var}
    }
    \caption{Variation of mean von Mises strain of bottom $W$ particle as a function of time following impacts at different velocities in the absence and presence of ultrasonic perturbation with the amplitude of $A=3.165\si{\angstrom}$ and frequency, $f=10$ GHz.}
    \label{fig:vel_variation_str_flat}
\end{figure}

As mentioned above, the maximum von Mises strain, $\epsilon_{mean}$, is considered as the parameter describing the deformation measure. \fref{fig:mean_str_nu_vel_var} shows the variation of \emean as a function of time. The saturation of \emean can be observed at each of the impact velocities in the absence of ultrasonic perturbation, which is in alignment with the previous observation of rate-independent plastic deformation shown in \fref{fig:800_no_ultra_mean_str}.  This saturation indicates that the material undergoes a finite amount of irreversible deformation and then reaches a stable state with no further strain accumulation, independent of the impact velocity. However, the maximum value of \emean increases as the impact velocity increases, which is a consequence of enhanced displacement of the atoms at higher impact velocities. A similar trend of \emean is also observed in ultrasound-assisted cases. However, the maximum value of \emean is nearly $1.5$ times higher than the values observed in non-ultrasound-assisted cases. The enhancement of plastic deformation is a consequence of acoustic softening and higher impact temperature caused by ultrasonic perturbation in the $W$ particles. While plastic deformation is enhanced in the ultrasound-assisted case, \emean saturates within a similar time interval as in the non-ultrasound cases.

\begin{figure}[t!]
    \centering
    \subfloat[No ultrasound-$1200$ms$^{-1}$]
    {
\includegraphics[width=0.5\textwidth]{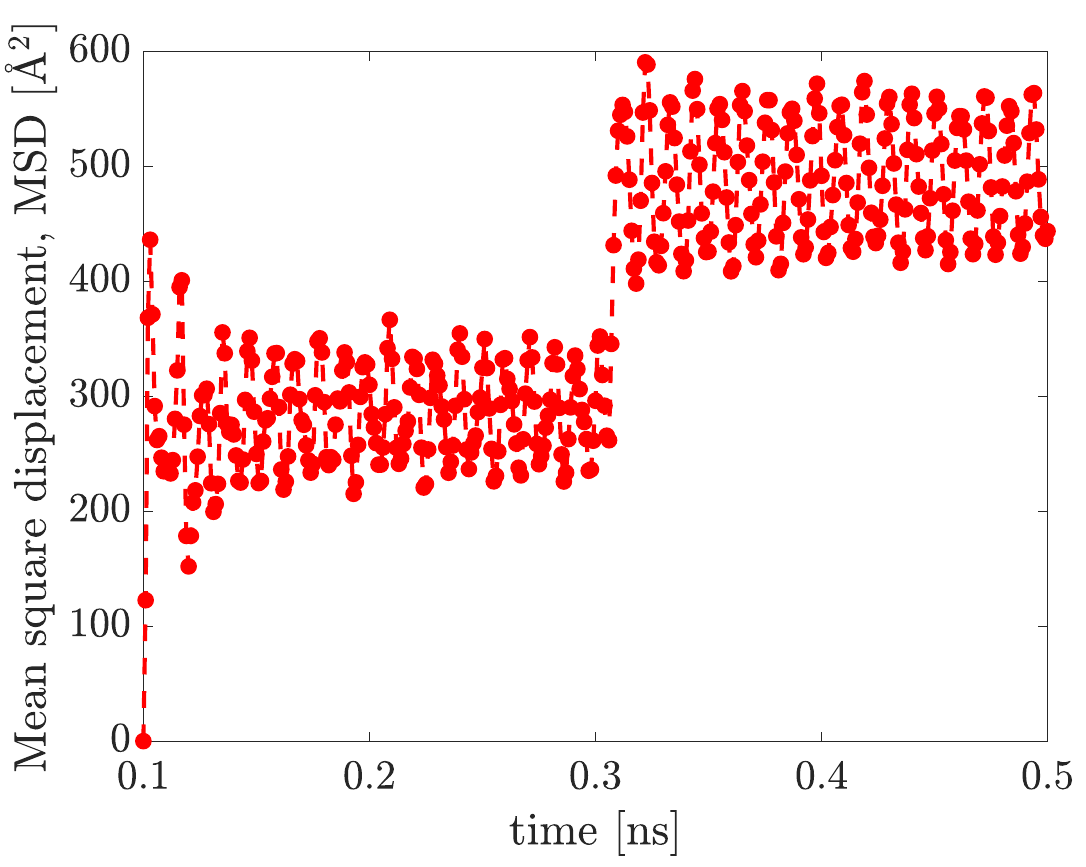}
    \label{fig:MSD_n_ultra_1200_c}
    }
    \subfloat[Ultrasound-$1200$ms$^{-1}$]
    {
\includegraphics[width=0.5\textwidth]{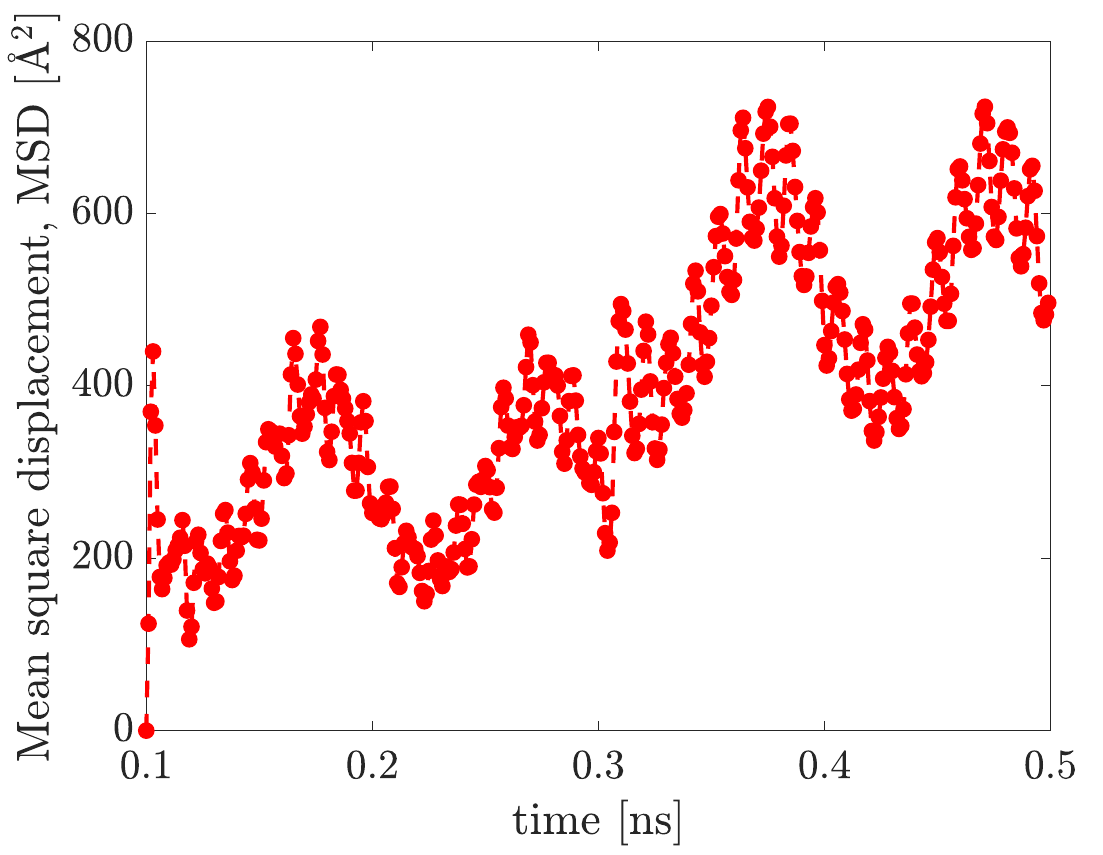}
    \label{fig:MSD_ultra_1200_c}
    }
    \caption{Variation of mean square displacement plot of bottom $W$ particle as a function of time at an impact velocity of $1200$ \ms \protect\subref{fig:MSD_n_ultra_1200_c} in the absence and, \protect\subref{fig:MSD_ultra_1200_c} presence of ultrasonic perturbation with the amplitude of $A=3.165\si{\angstrom}$ and frequency, $f=10$ GHz.}
    \label{fig:MSD_plots_ultra}
\end{figure}

While the evolution of mean von Mises strain, \emean, effectively captures the deformation behavior, it does not provide insight into the underlying atomic mobility. To address this, we computed the MSD evolution of atoms within the bottom particle at various impact velocities, with and without ultrasound.\fref{fig:800_no_ultra_MSD}, and \fref{fig:MSD_n_ultra_1200_c} compare the evolution of MSD as a function of time at different impact velocities. We note that MSD exhibits an oscillatory profile which, however, follows a steady mean. Such a steady mean indicates limited or no atomic diffusion post-impact, irrespective of impact velocity. In contrast, ultrasound-assisted configurations exhibit an ultrasound-frequency-dependent MSD plot, as illustrated in \fref{fig:800_ultra_MSD} and \fref{fig:MSD_ultra_1200_c}. While the frequency of the MSD plot is governed by the frequency of the imposed ultrasonic perturbation, the overall slope of these plots remains zero, indicating the absence of a diffusive nature. This particular observation is commensurate with the fundamentals of the CS process, where diffusion is negligible without any addition of external heat treatment \cite{yin2024towards}. Thus, it can be asserted that self-diffusion in CS of W is absent irrespective of the applied impact velocities.  

\begin{figure}[t!]
    \centering
    \subfloat[RDF-No Ultrasound]
    {
\includegraphics[width=0.5\textwidth]{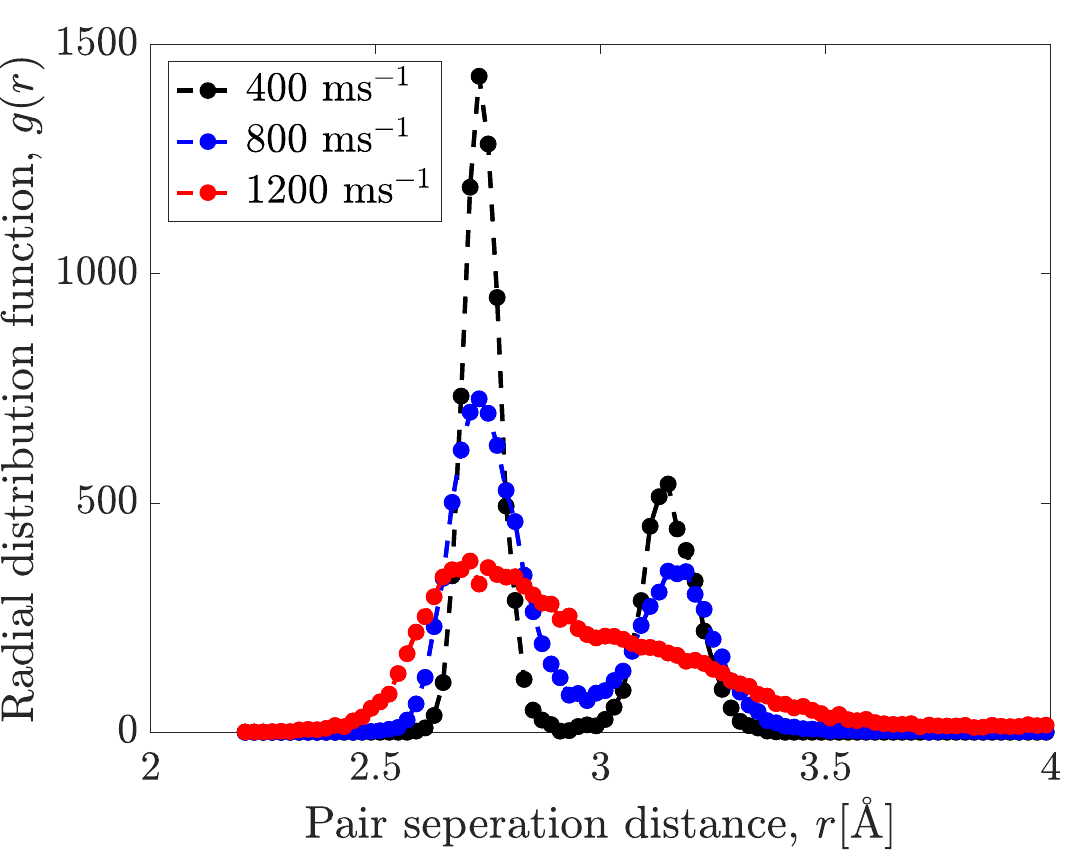}
    \label{fig:rdf_val_ver_nu}
    }
    \subfloat[RDF-Ultrasound]
    {
\includegraphics[width=0.5\textwidth]{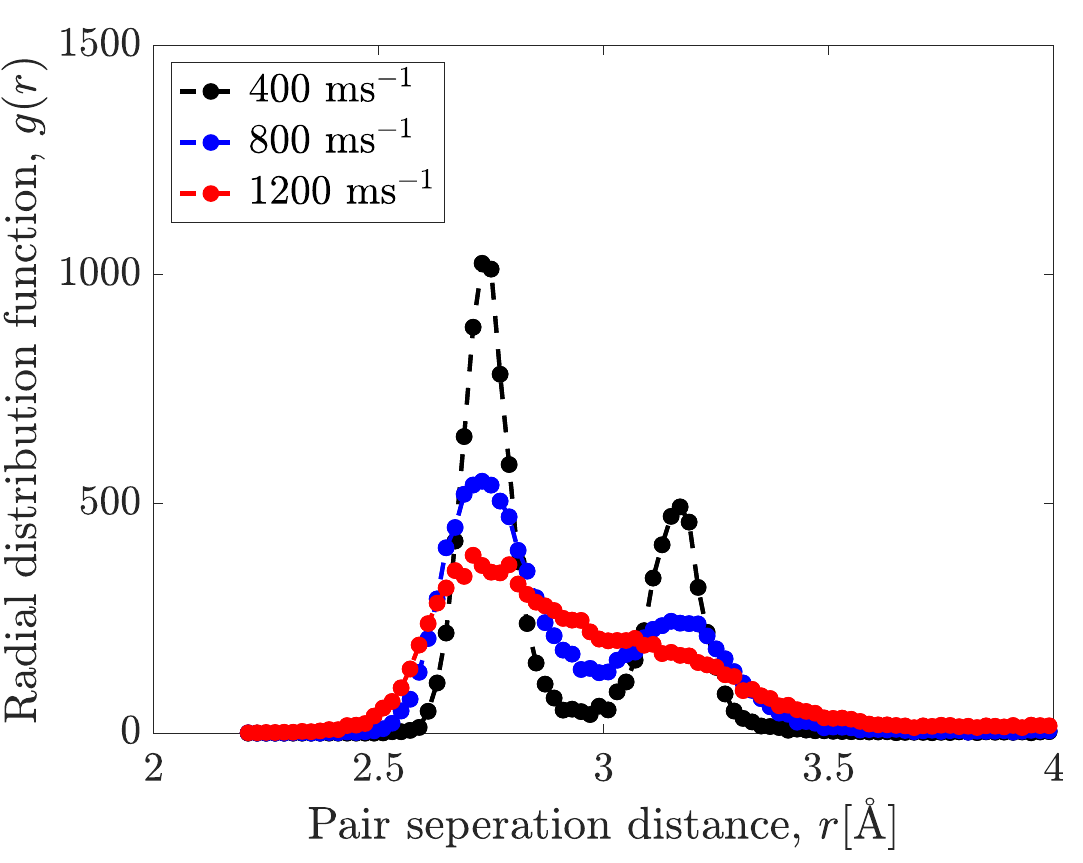}
    \label{fig:rdf_val_ver_u}
    }
    \caption{Variation of radial distribution function of bottom $W$ particle at time, $t=0.11$ns following impacts at different velocities in the absence and presence of ultrasonic perturbation with the amplitude of $A=3.165\si{\angstrom}$ and frequency, $f=10$ GHz.}
    \label{fig:vel_variation_rdf}
\end{figure}

To connect the velocity-dependent deformation behavior to long-range order in the atomic configurations, we compute the RDF at $t=0.11$ ns at different impact velocities. As shown in \fref{fig:vel_variation_rdf}, for both ultrasound-assisted and non-ultrasound-assisted cases, crystallinity can be observed at $400$ \ms and $800$ \ms impact velocities. However, the width of the RDF peaks is larger in ultrasound-assisted cases, indicating considerable plastic deformation. However, at $1200$ \ms, the RDF distributions correspond to an amorphous long-range order due to a high temperature surge caused by impact at such a high impact velocity. However, this amorphous condition is temperature-dependent and does not persist for long after the impact, as similarly shown in \fref{fig:temp_track} for $800$ \ms.  

\begin{figure}
    \centering
    \includegraphics[width=0.5\linewidth]{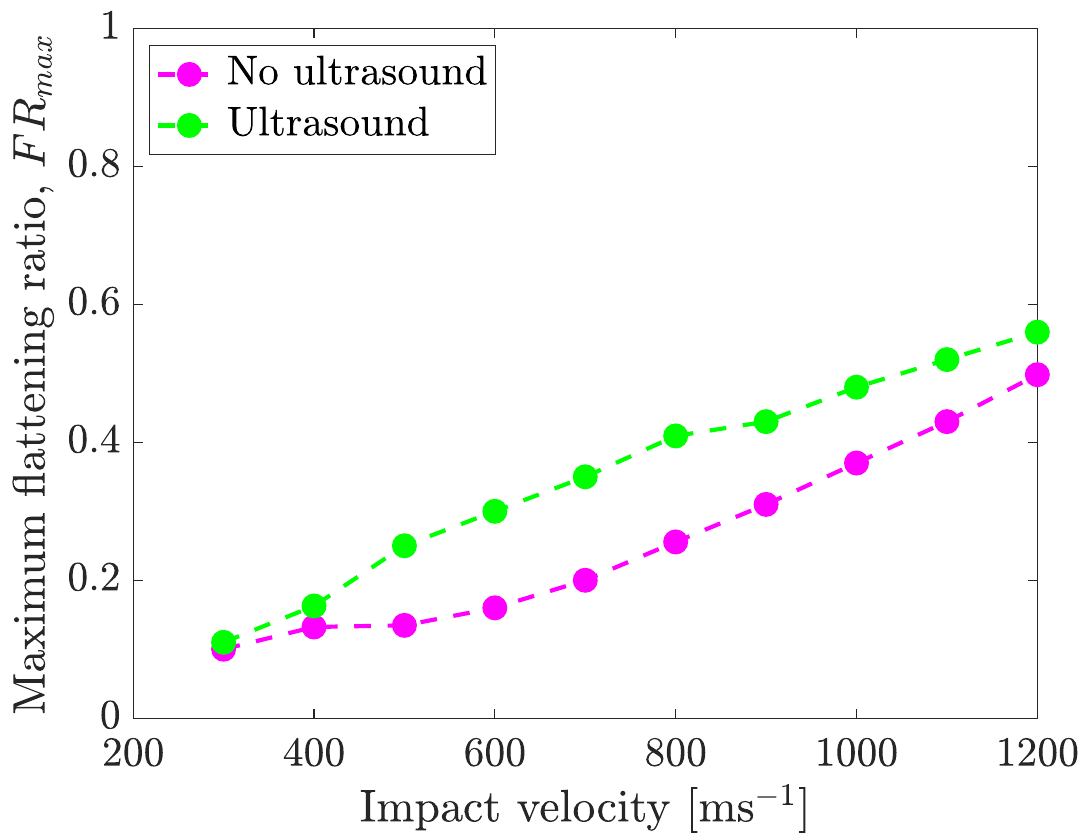}
    \caption{Variation of maximum flattening ratio of bottom $W$ particle as a function of impact velocity in the absence and presence of ultrasonic perturbation with the amplitude of $A=3.165\si{\angstrom}$ and frequency, $f=10$ GHz.}
    \label{fig:FR_u}
\end{figure}

Finally, to illustrate the effect of impact velocities on the deposited coating quality, we compute the flattening ratio, $FR_{max}$, at different impact velocities in the absence and presence of ultrasonic perturbation. \fref{fig:FR_u} compares the maximum flattening ratio under these two conditions. It can be observed that $FR_{max}$ attains a maximum value of $0.5$ in the absence of ultrasonic perturbation. Moreover, in the absence of ultrasonic perturbation and at low impact velocities of $300-500$ \ms, $FR_{max}$ does not strongly respond to the variation of impact velocities. Above $500$ \ms, $FR_{max}$ keeps on increasing linearly. This can be explained by the insignificant plastic deformation observed at lower velocities in the absence of ultrasonic perturbation. In contrast, $FR_{max}$ attains a maximum value of $0.56$ in the presence of an ultrasonic perturbation with an amplitude of $A=3.165\si{\angstrom}$ and a frequency of $f=10$ GHz. A consistent linear increase is observed in the ultrasound-assisted case, indicating that flattening is driven not only by the impact but also by acoustic softening in ultrasound-assisted CS. Thus, we observe that ultrasound promotes plastic deformation of $W$, leading to formation of a uniform coating during CS process. While the variation in impact velocity demonstrates the generality of ultrasound-assisted acoustoplasticity in CS of $W$, we explore its dependence on particle size in the next section.

\subsection{Size of the particle}
\label{sec:part_size}
In the previous section, we presented how impact velocities affect the plastic deformation of $W$ particles during the CS process. In this section, we present whether particle diameter affects the deformation mechanism and atomic rearrangements of $W$ particles during the CS process. 

\begin{figure}[t!]
    \centering
    \subfloat[$D=37.98\si{\angstrom}$-No Ultrasound]
    {
\includegraphics[width=0.28\textwidth]{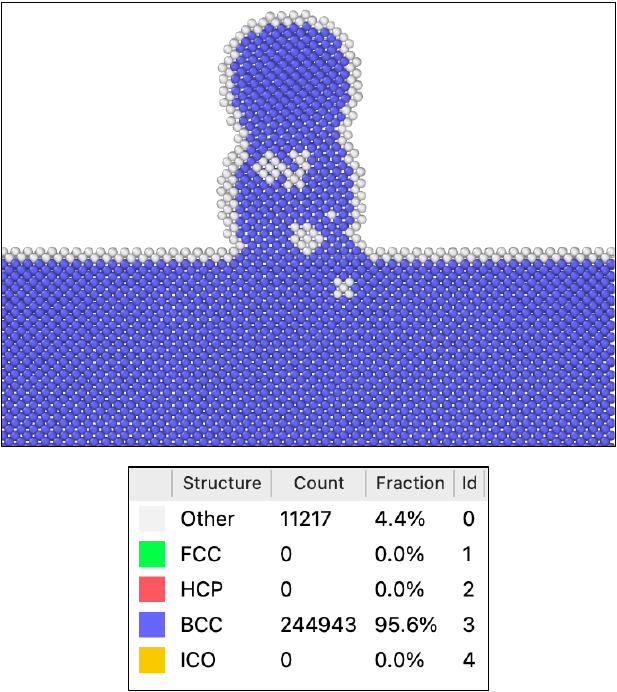}
    \label{fig:config_6_800}
    }
    \subfloat[$D=63.30\si{\angstrom}$-No Ultrasound]
    {
\includegraphics[width=0.28\columnwidth]{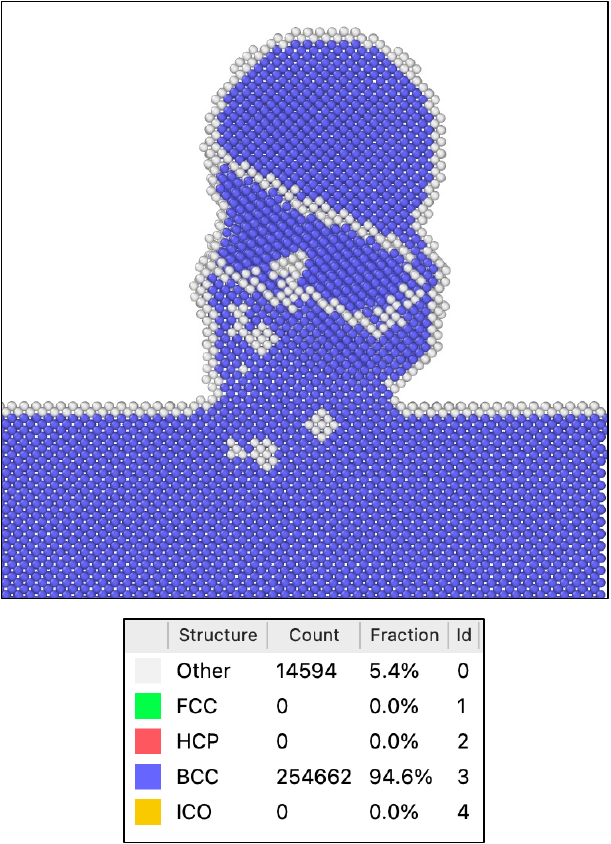}
        \label{fig:config_10_800}
    } 
        \subfloat[$D=75.96\si{\angstrom}$-No Ultrasound]
    {
\includegraphics[width=0.28\columnwidth]{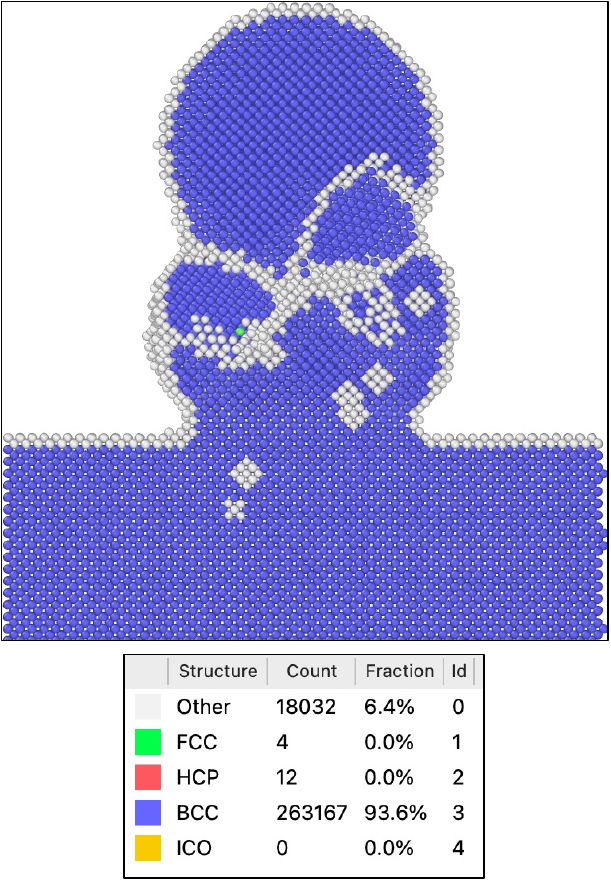}
        \label{fig:config_12_800}
    } 
    \\
    \subfloat[$D=37.98\si{\angstrom}$-Ultrasound]
    {
\includegraphics[width=0.28\textwidth]{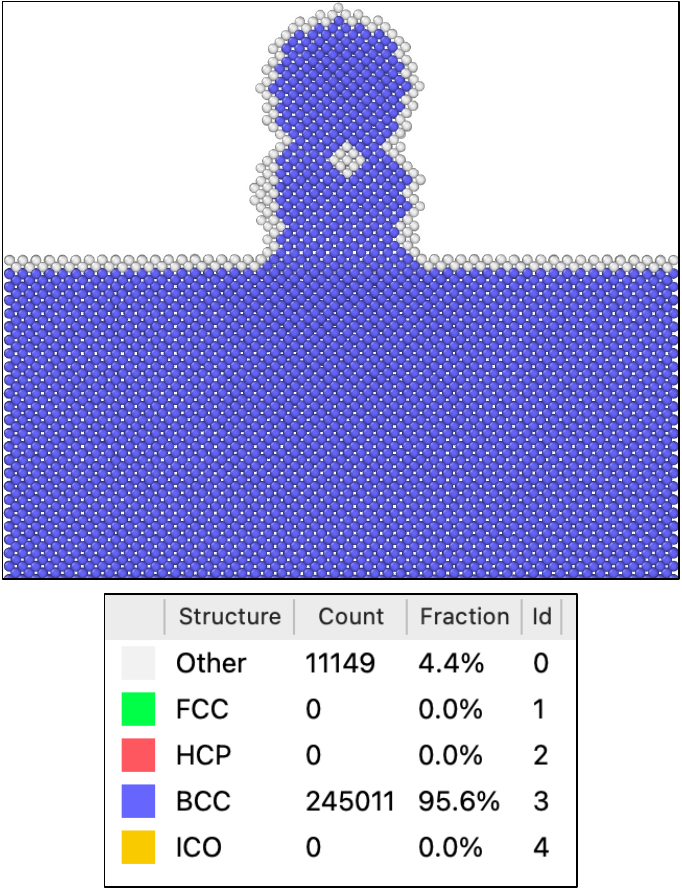}
    \label{fig:config_6_u_800}
    }
    \subfloat[$D=63.30\si{\angstrom}$-Ultrasound]
    {
\includegraphics[width=0.28\columnwidth]{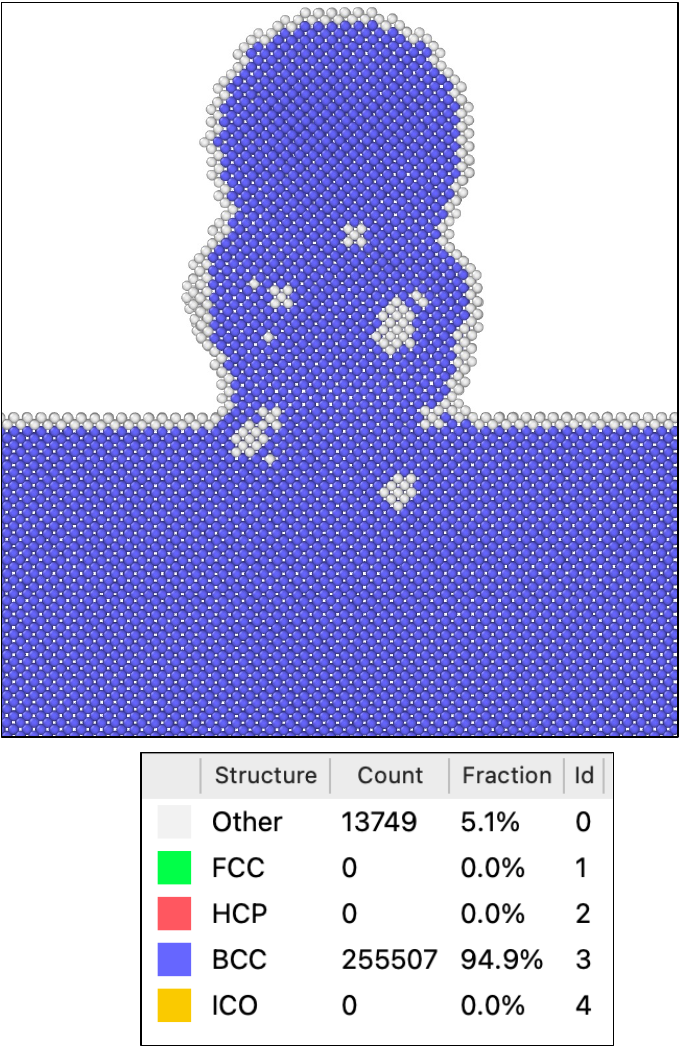}
        \label{fig:config_10_u_800}
    } 
        \subfloat[$D=75.96\si{\angstrom}$-Ultrasound]
    {
\includegraphics[width=0.28\columnwidth]{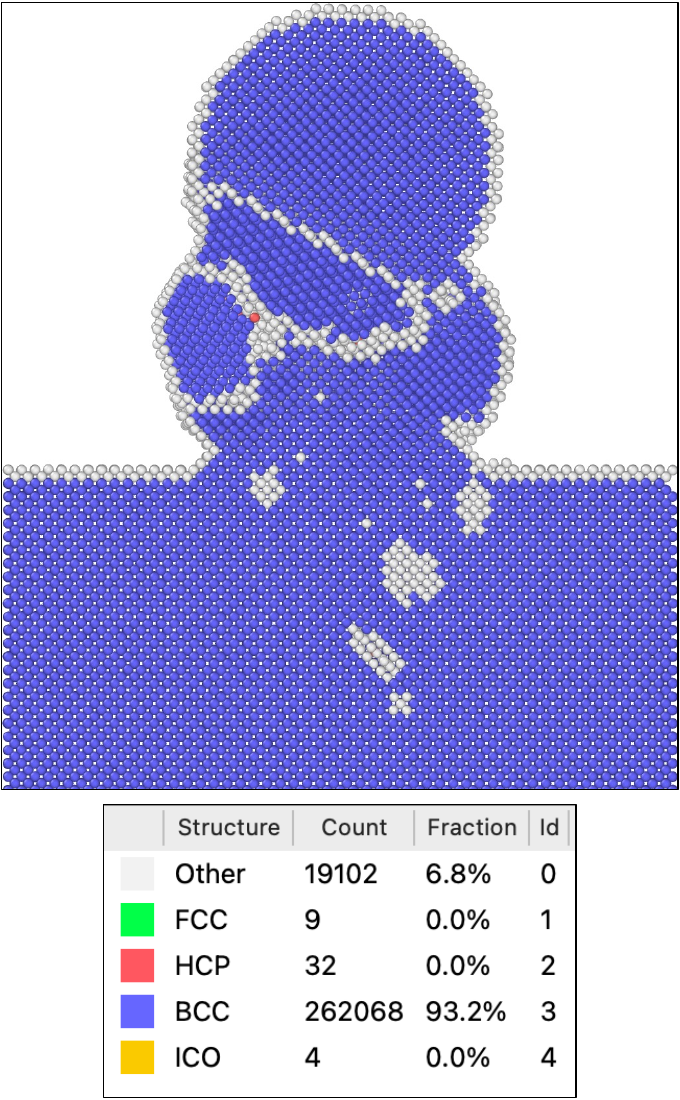}
        \label{fig:config_12_u_800}
    } 
    \caption{Variation of atomic configurations of $W$ particles and substrate as a function of particle diameters following impact at an impact velocity of $800$ ms$^{-1}$ in the absence [\protect\subref{fig:config_6_800}, \protect\subref{fig:config_10_800} and \protect\subref{fig:config_12_800}] and presence [\protect\subref{fig:config_6_u_800}, \protect\subref{fig:config_10_u_800} and \protect\subref{fig:config_12_u_800}] of ultrasonic perturbation with the amplitude of $A=3.165\si{\angstrom}$ and frequency, $f=10$ GHz at time $t=0.5$ ns. The number of different lattice structures corresponding to each atomic configuration is shown below the figures.}
    \label{fig:config_plots_nu_u_800}
\end{figure}

First, we present the variation in atomic configurations as the $W$ particle diameter is varied from $37.98\si{\angstrom}$ to $75.96\si{\angstrom}$ for both the non-ultrasound and ultrasound-assisted CS processes. \fref{fig:config_6_800}, \fref{fig:800_no_ultra_cal}, \fref{fig:config_10_800}, and \fref{fig:config_12_800} show the variation of atomic configurations for different sizes of the $W$ particles when the particles are subjected to $800$ \ms in the absence of ultrasonic perturbation. We note pore formation is the only defect observed in small particles ($D \le 50.64\si{\angstrom}$) while large particles form both pore and grain boundaries during the impact. The grain boundary formation in large particles is a consequence of the larger surface boundaries, which merge to form long grain boundaries. The grain boundaries resist the plastic deformation, impeding the formation of an expanded coating on the $W$ substrate. Moreover, the $W$ particles only pile up one on top of another and cause instability of the deposited coating in the absence of ultrasonic perturbation at low velocities. However, in the presence of ultrasonic perturbation, the particles undergo constant agitation, leading to dissolution of the larger grain boundaries and subsequent grain refinement. This leads to a decrease in the overall number of pores in the ultrasound-assisted case. \fref{fig:config_6_u_800}, \fref{fig:800_no_ultra_cal}, \fref{fig:config_10_u_800}, and \fref{fig:config_12_u_800} highlight the variation of atomic configurations as a function of particle diameters for the ultrasound-assisted case. As mentioned before,  ultrasound promotes grain refinement and recrystallization, leading to a reduction in non-crystalline interfaces for $W$ particles with a diameter smaller than $D<75.96\si{\angstrom}$. At larger particles, the lower percentage of BCC fraction can be attributed to the insufficient time available for grain refinement of larger particles.

\begin{figure}[t!]
    \centering
    \subfloat[]
    {
\includegraphics[width=0.5\textwidth]{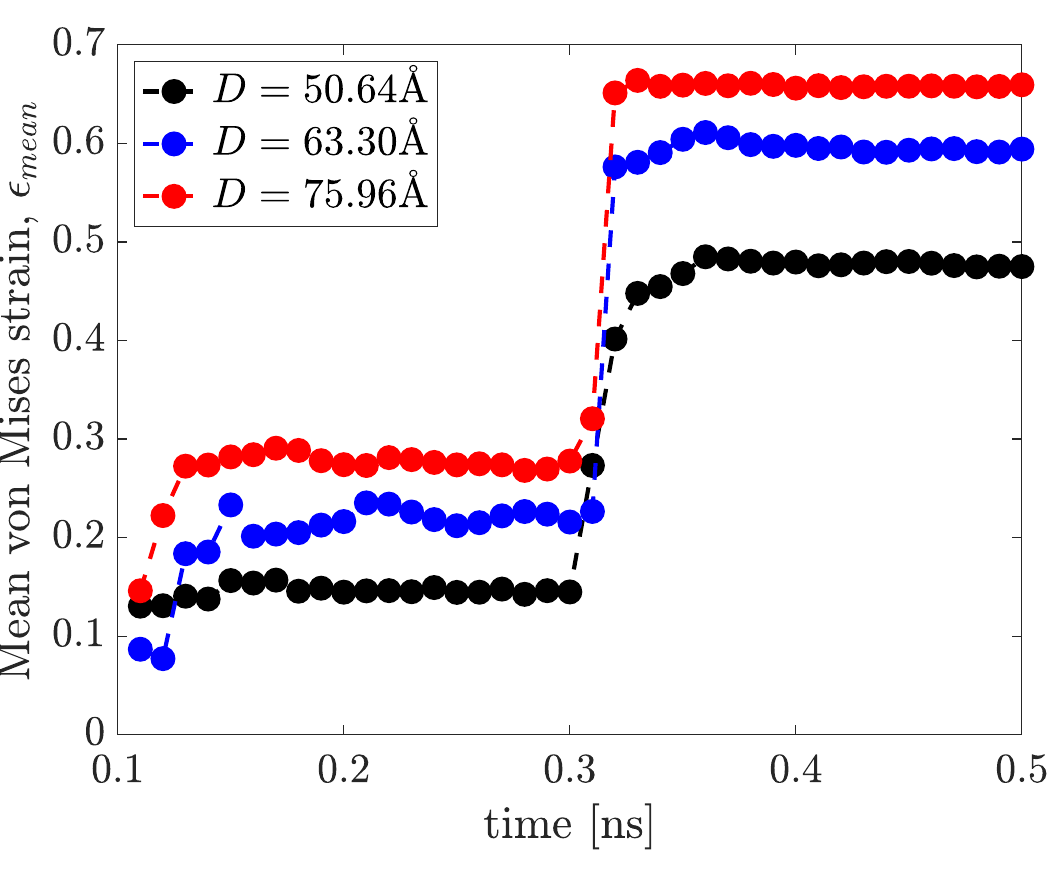}
    \label{fig:mean_str_n_ultra_800}
    }
    \subfloat[]
    {
\includegraphics[width=0.5\columnwidth]{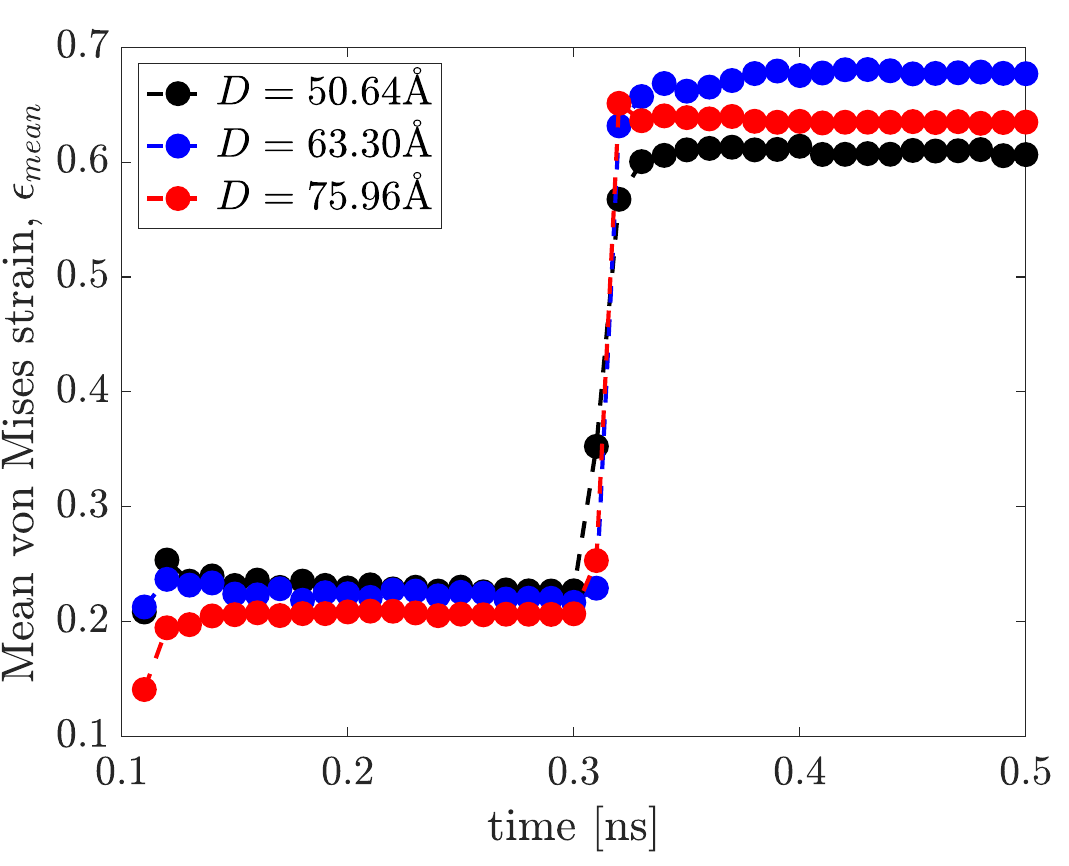}
        \label{fig:mean_str_ultra_800}
    } 
    \caption{Variation of mean strain of the bottom particle, $\epsilon_{mean}$ for impact velocity of $800$ ms$^{-1}$ as a function of time for different particle sizes. \protect\subref{fig:mean_str_n_ultra_800} and \protect\subref{fig:mean_str_ultra_800} compare $\epsilon_{mean}$ without and with the application of ultrasound respectively.
    }
    \label{fig:mean_plots_ultra_800}
\end{figure}

To quantify the contribution of particle diameters to deformation during the CS process, we compute the evolution of \emean for the above-mentioned ranges of particle diameters. \fref{fig:mean_str_n_ultra_800} shows the variation of \emean as a function of time for three different particle diameters under the non-ultrasound-assisted condition. In large particles, atoms farther from the center are subjected to greater deformation, leading to greater local strain accumulation. However, the formation of grain boundaries resists plastic deformation, counterbalancing this effect. This non-linear increment might lead to a size-insensitive plastic deformation as the particle diameter reaches a critical maximum. However, the similar behavior is not apparent in cases of ultrasound-assisted cases. A saturation of the \emean profile is observed across all particle diameters. As observed in \fref{fig:mean_str_ultra_800}, the plastic deformation is relatively insensitive to the changes in particle diameters, which leads to hypothesize a strong dependence of ultrasound parameters and impact velocity in governing plastic deformation in ultrasound-assisted CS rather than the size of the particle itself.

\begin{figure}[t!]
    \centering
    \subfloat[]
    {
\includegraphics[width=0.5\textwidth]{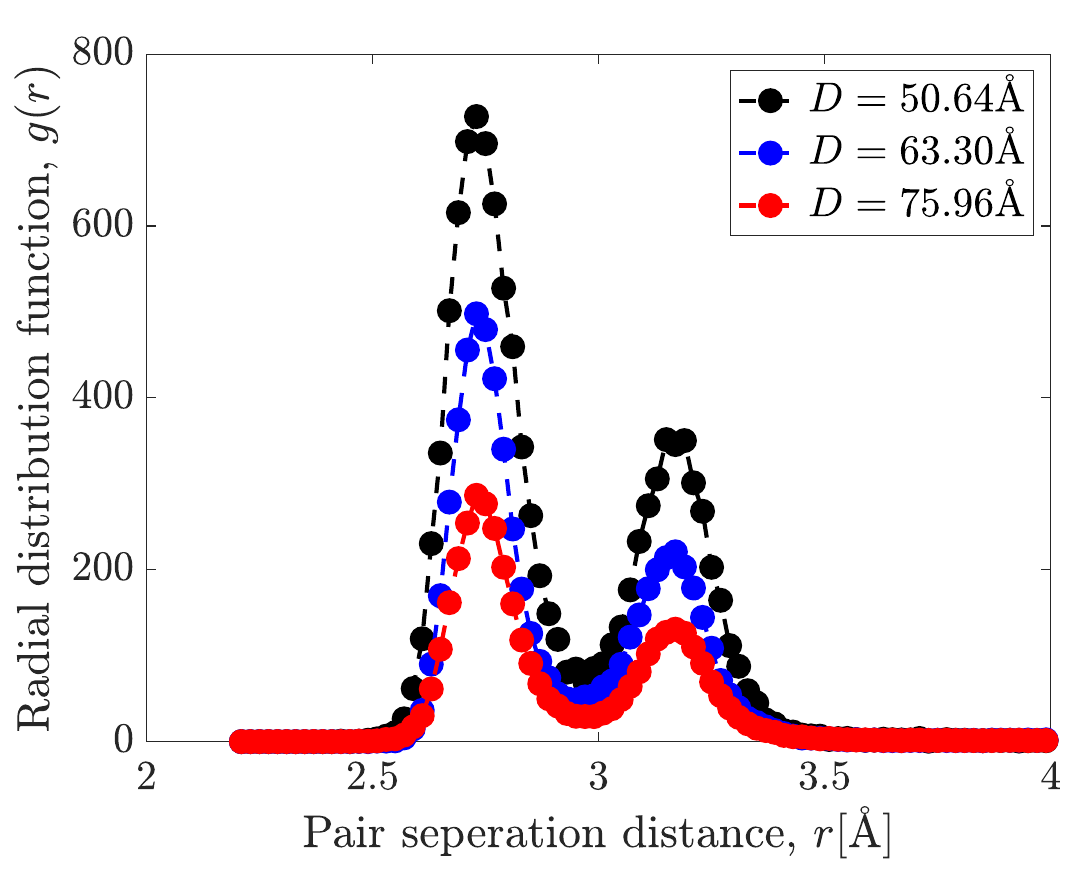}
    \label{fig:rdf_n_ultra_800}
    }
    \subfloat[]
    {
\includegraphics[width=0.5\columnwidth]{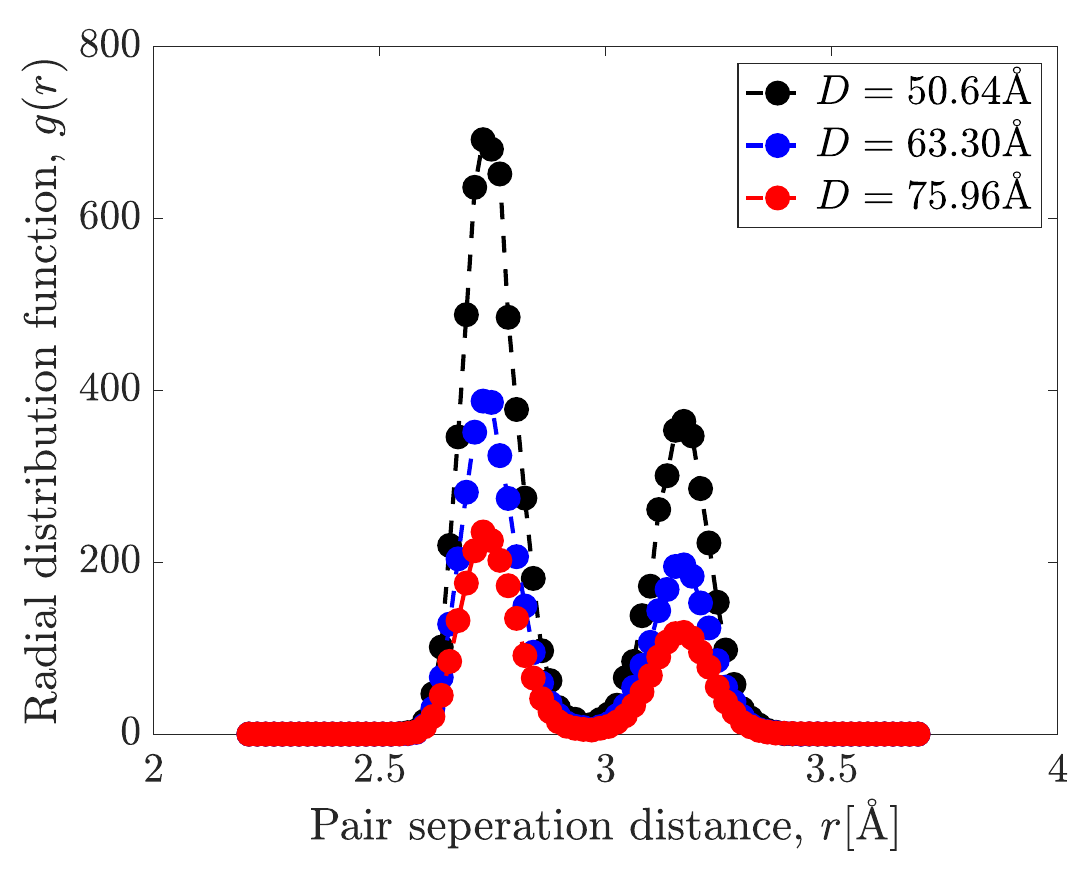}
        \label{fig:rdf_ultra_800}
    } 
    \caption{Variation of radial distribution function for impact velocity of $800$ ms$^{-1}$ at $t=0.11$ ns \protect\subref{fig:rdf_n_ultra_800} without and, \protect\subref{fig:rdf_ultra_800} with the application of ultrasound.}
    \label{fig:rdf_plots_ultra_800}
\end{figure}

To correlate the deformation mechanism to the long-range order of $W$ particles, we computed the RDF distribution for different sets of particle diameters at time $t=0.11$ ns in the absence and presence of the ultrasonic perturbation. \fref{fig:rdf_n_ultra_800} shows the RDFs for three particle diameters, which illustrate the crystalline nature of the bottom particle after impact with the substrate and of the top particle. Similar to the non-ultrasound-assisted case, \fref{fig:rdf_ultra_800} shows the crystalline RDF distribution. The reduction in peak height as the particle size increases is due to the enhanced local disorder observable in larger particles. 

\begin{figure}
    \centering
    \includegraphics[width=0.5\linewidth]{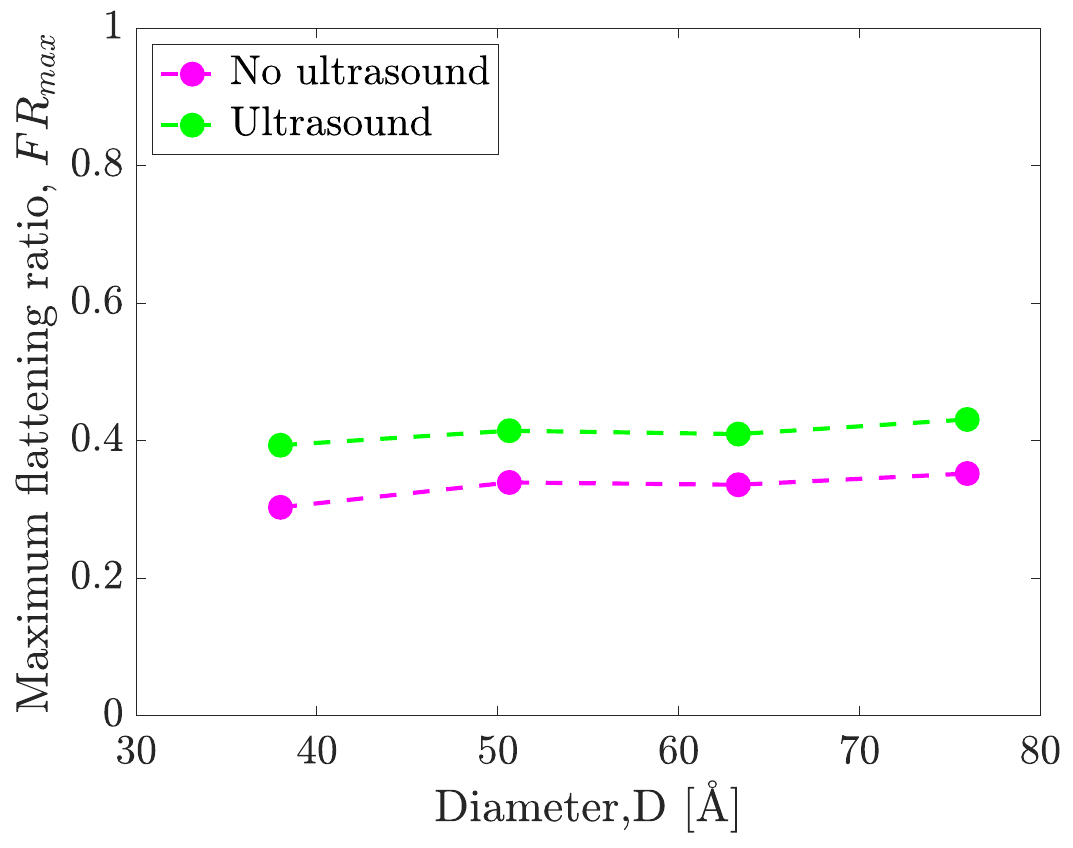}
    \caption{Variation of maximum flattening ratio of bottom $W$ particle as a function of particle diameters in the absence and presence of ultrasonic perturbation with the amplitude of $A=3.165\si{\angstrom}$ and frequency, $f=10$ GHz.}
    \label{fig:FR_u_size}
\end{figure}

In the previous sections, we investigated how particle dimensions affect the deformation and atomic rearrangements of $W$ particles in the CS process. Finally, we present how particle size affects the quality of the deposited $W$ coating. \fref{fig:FR_u_size} shows the variation of $FR_{max}$ as a function of particle diameter in the absence and presence of ultrasonic perturbation. We note that both without and with ultrasonic perturbation, $FR_{max}$ does not change considerably with the particle diameter. While particle size might cause localized deformation in larger particles, leading to a higher \emean in the non-ultrasound-assisted case, the average ratio of deformed height to initial diameter remains the same. However, ultrasound parameters can affect the deformation measure and atomic rearrangements in the ultrasound-assisted CS process. We explore this in the next section.

\subsection{Frequency effect}
\label{sec:freq_eff}
In the previous sections, we demonstrated that impact velocity affects the deformation and rearrangement rates of the atomic configurations of $W$ particles in both ultrasound and non-ultrasound-assisted CS processes. We also showed that plastic deformation is nearly insensitive to particle size variation in ultrasound-assisted CS. In this section, we illustrate how ultrasound parameters, amplitude and frequency, affect the deformation behavior and atomic rearrangements of $W$ particles during the CS process.  

\begin{figure}[t!]
    \centering
    \subfloat[]
    {
\includegraphics[width=0.5\columnwidth]{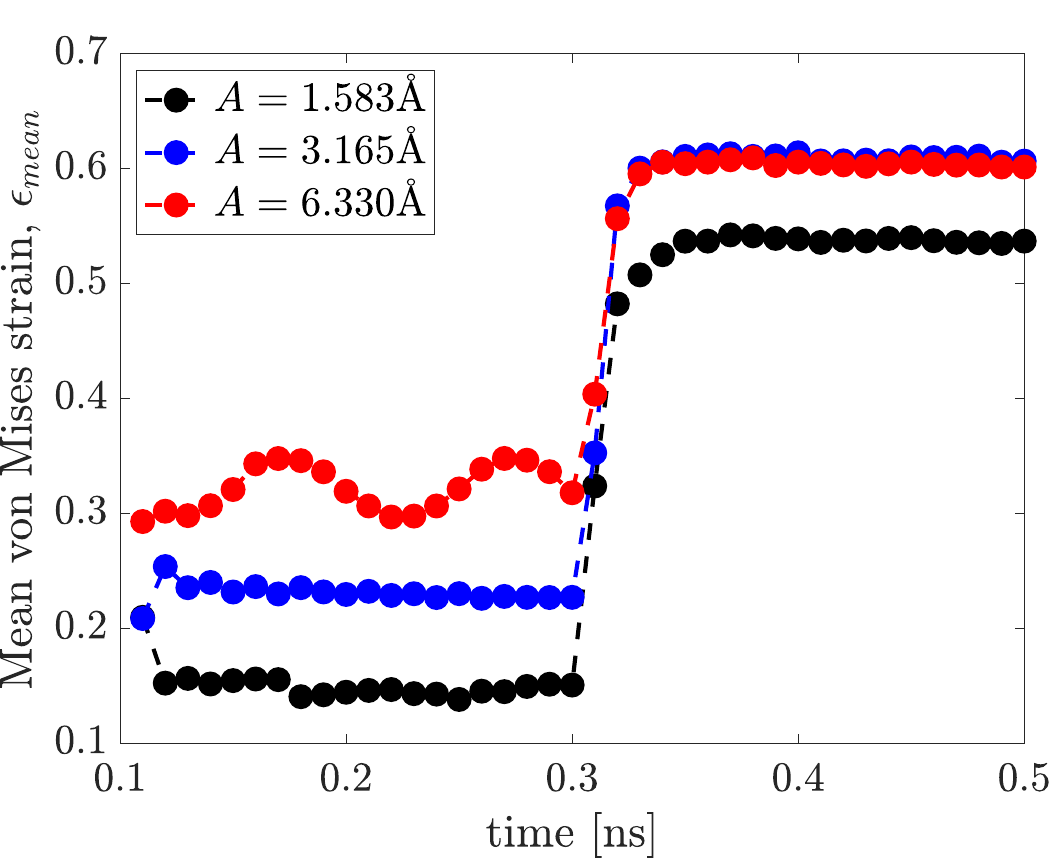}
        \label{fig:mean_var_amp}
    } 
    \subfloat[]
    {
\includegraphics[width=0.5\columnwidth]{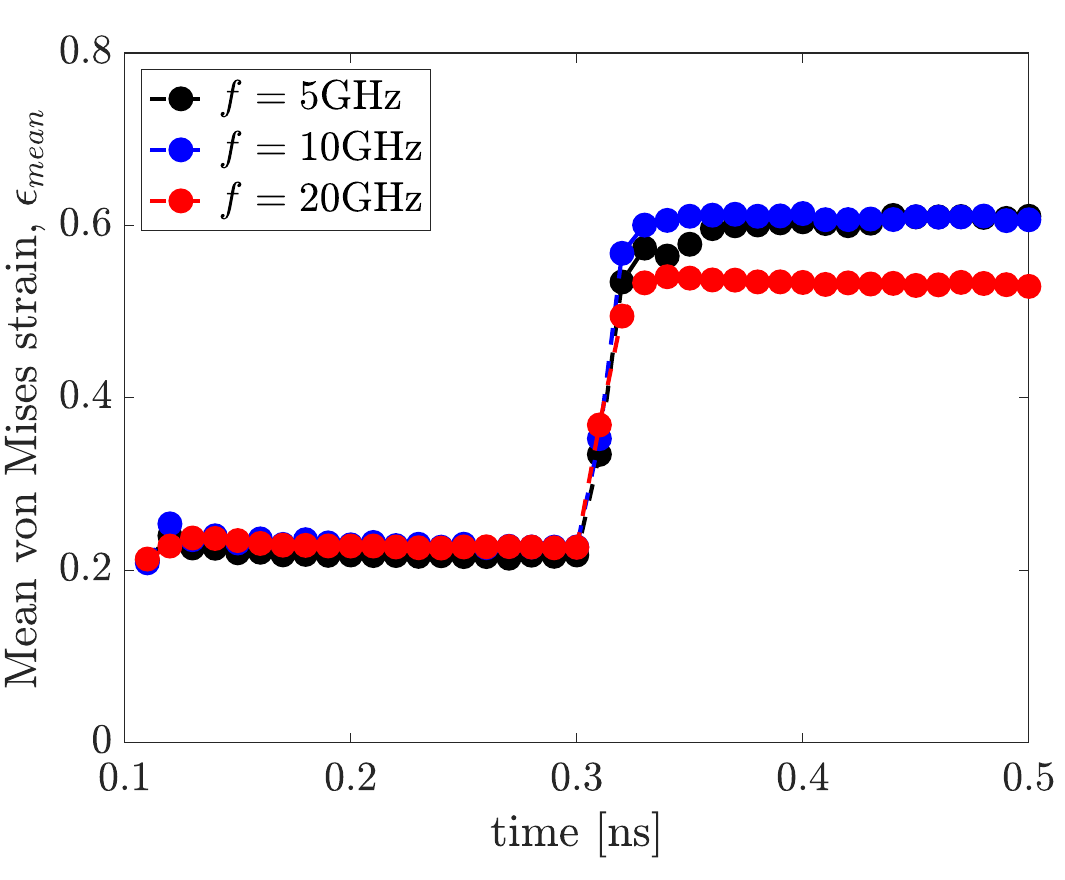}
        \label{fig:mean_var_freq}
    }
    \caption{Variation of mean von Mises strain as a function of time for impact velocity of $800$ \ms at different ultrasound \protect\subref{fig:mean_var_amp} amplitudes and \protect\subref{fig:mean_var_freq} frequencies.
    }
    \label{fig:diff_freq_str}
\end{figure}

First, we present the effects of ultrasound amplitude and frequency on the evolution of \emean over time. \fref{fig:mean_var_amp} shows the evolution of \emean as a function of time for different amplitudes of ultrasonic perturbation. We note that the maximum value of \emean saturates as the amplitude increases from $A=3.165\si{\angstrom}$ to $A=6.33\si{\angstrom}$. Such a saturation is governed by the maximum flattening attainable in the presence of an ultrasonic perturbation.  However, increasing the ultrasound frequency has the opposite effect. \fref{fig:mean_var_freq} shows that the maximum value of \emean increases $12\%$ as the frequency is lowered from $f=20$GHz to $f=10$GHz. However, up to $t=0.3$ns, the evolution of the \emean plot converges for all the frequencies. Such convergences at higher amplitudes and lower frequencies set a limit on the maximum attainable plastic deformation in ultrasound-assisted CS processes.     

\begin{figure}[t!]
    \centering
            \subfloat[]
    {
\includegraphics[width=0.5\columnwidth]{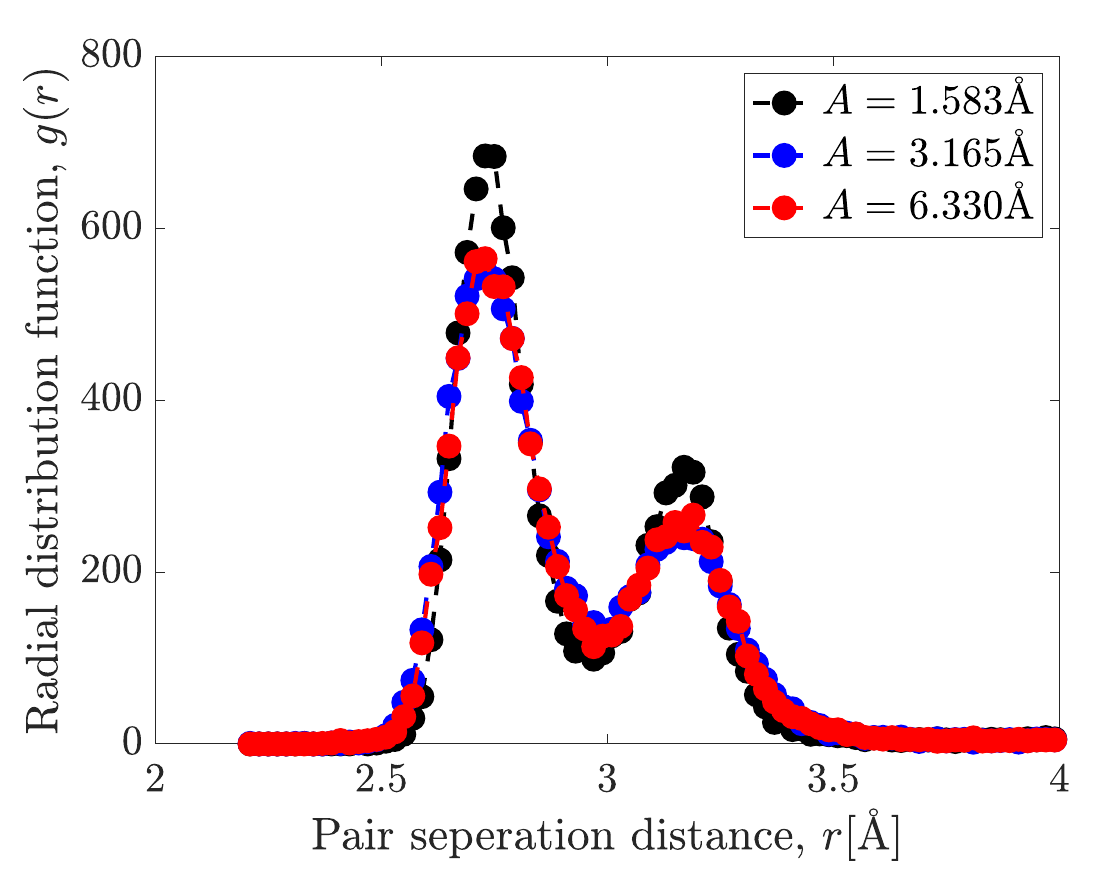}
        \label{fig:gr_amp}
    } 
    \subfloat[]
    {
\includegraphics[width=0.5\columnwidth]{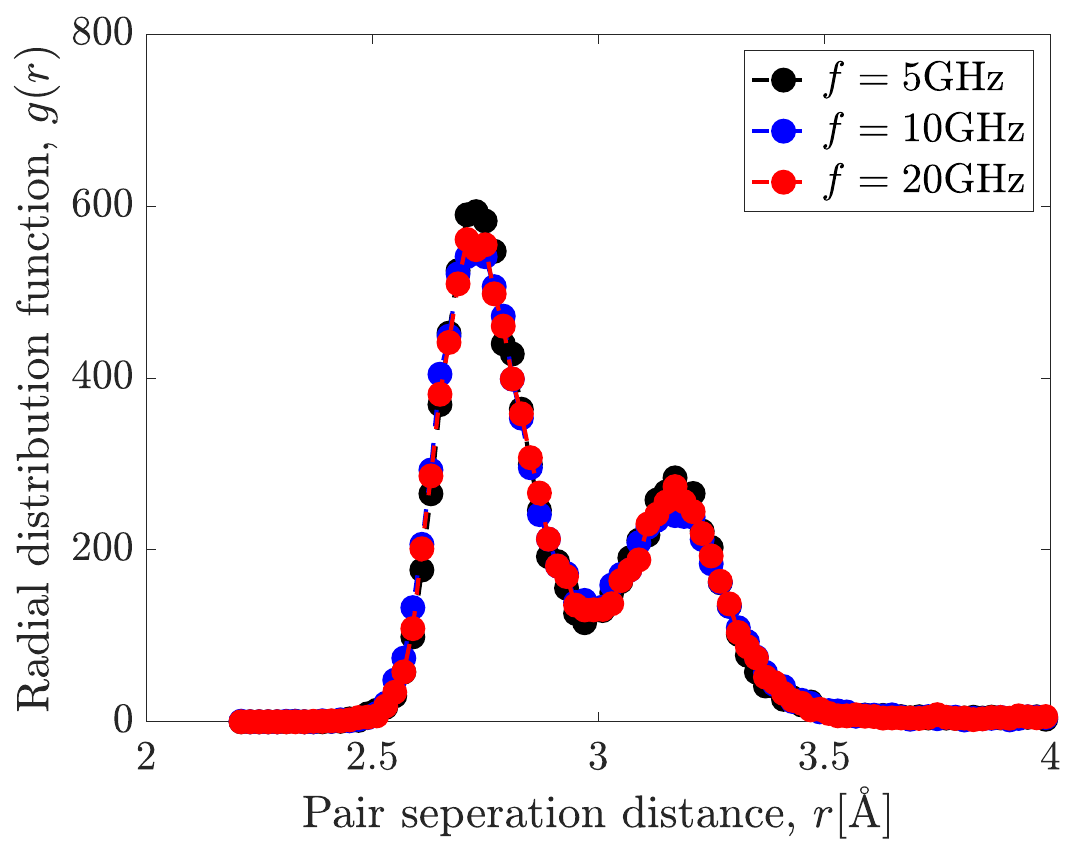}
        \label{fig:gr_freq}
    }
    \caption{Variation of radial distribution function for impact velocity of $800$ ms$^{-1}$ at $t=0.11$ ns as a function of ultrasonic perturbation \protect\subref{fig:gr_amp} amplitude and, \protect\subref{fig:gr_freq} frequency.
    }
    \label{fig:diff_freq_gr}
\end{figure}

To characterize the effect of ultrasound parameters on the long-range order of $W$ particles, we compute the RDFs for the bottom particle of $W$ for the ultrasound-assisted case at time $t=0.11$ns, where the $W$ substrate is subjected to different ultrasound amplitudes and frequencies. As observed in \fref{fig:gr_freq}, the ultrasound frequency does not modify the RDF distributions. This suggests that variations in ultrasound frequency enhance local yielding by promoting the displacement of energetically favorable atoms. At the same time, the overall long-range atomic arrangement is preserved throughout the CS process for a given impact velocity and particle diameter. Similar to the variation of frequencies, \fref{fig:gr_amp} shows similar RDF distributions between the amplitudes of $A=3.165-6.33\si{\angstrom}$. However, at a low amplitude of $A=1.583\si{\angstrom}$, the particle undergoes less plastic deformation, as also shown in \fref{fig:mean_var_amp}, which leads to the higher peak of the RDF distribution. 

\begin{figure}[t!]
    \centering
    \subfloat[]
    {
\includegraphics[width=0.5\textwidth]{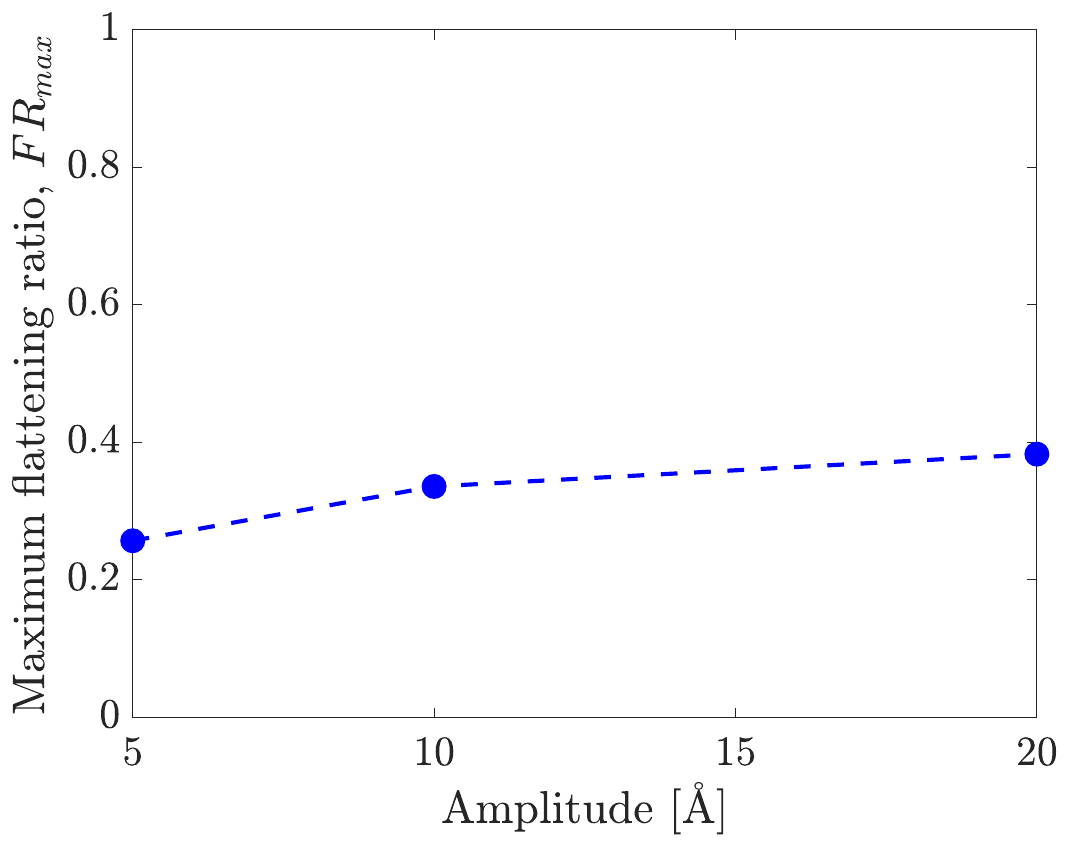}
    \label{fig:flatten_amp}
    }
    \subfloat[]
    {
\includegraphics[width=0.5\columnwidth]{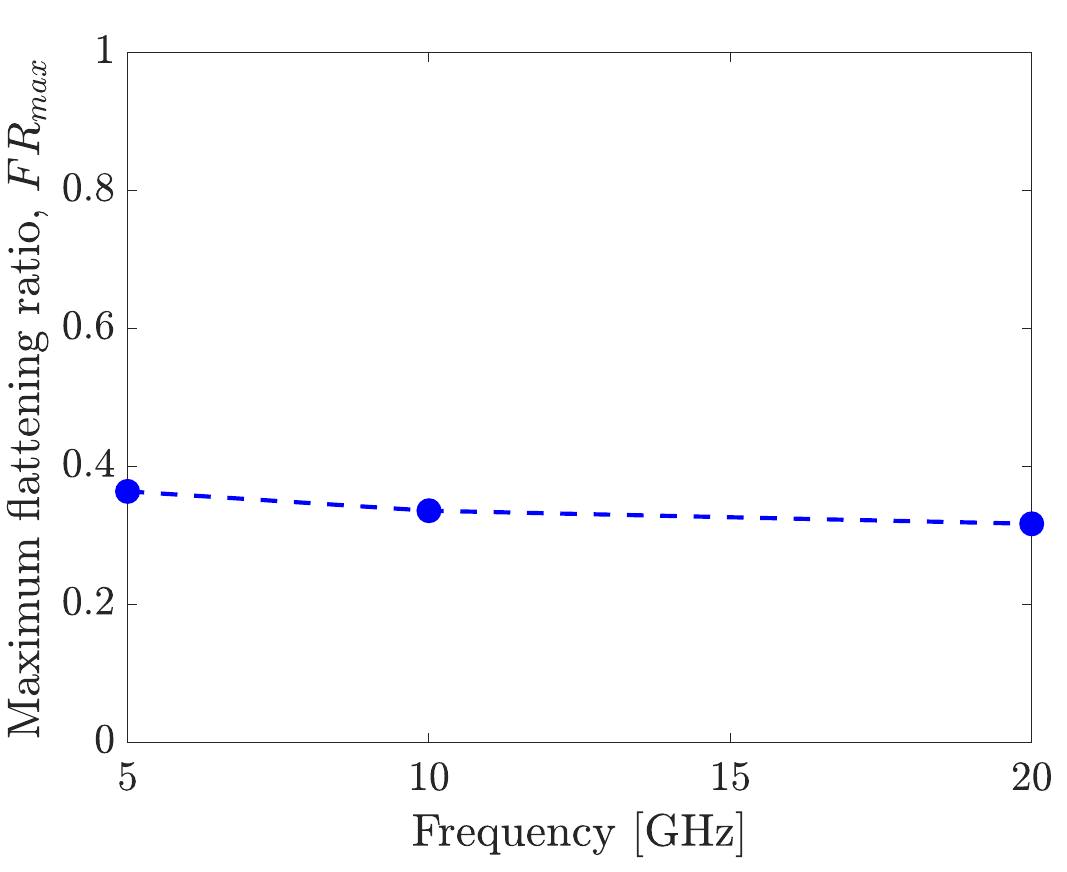}
        \label{fig:flatten_freq}
    } 
    \caption{Variation of maximum flattening ratio for impact velocity of $800$ ms$^{-1}$ at $t=0.5$ ns as a function of ultrasonic perturbation \protect\subref{fig:flatten_amp} amplitude and \protect\subref{fig:flatten_freq} frequency.
    }
    \label{fig:diff_freq_fr}
\end{figure}

Similar consistency can be observed in the plots shown in \fref{fig:diff_freq_fr}. Here, the flattening ratio shows little response to variations in ultrasound parameters. Uniformity in the global atomic arrangement of the bottom $W$ particles appears to contribute to the saturation of parameters such as RDF and $FR_{max}$, while a substantial change in \emean occurs due to local yielding near the impact region. 

As shown in the previous sections, higher plastic deformation in $W$ can be achieved by applying ultrasonic perturbations during the CS process. Whether interatomic bonding between dissimilar elements occurs during the CS process, will be explored in the next section.  

\section{Acoustoplastic softening mediated cold spray of Vanadium -Tungsten alloy}
\label{sec:alloy_form}
In the previous sections, we presented that acoustoplasticity in the ultrasound-assisted CS process contributes to enhanced plastic deformation. While this is useful for pure metals, one of the more pronounced applications could be CSAM with prealloyed particles to form heterogeneous interface. In the next section, we present a case study of the formation of a $V$-$W$ alloy coating on a $W$ substrate using CS via ultrasonic perturbation. 

\begin{figure}[t!]
    \centering
    \subfloat[]
    {
    \includegraphics[width=0.5\columnwidth]{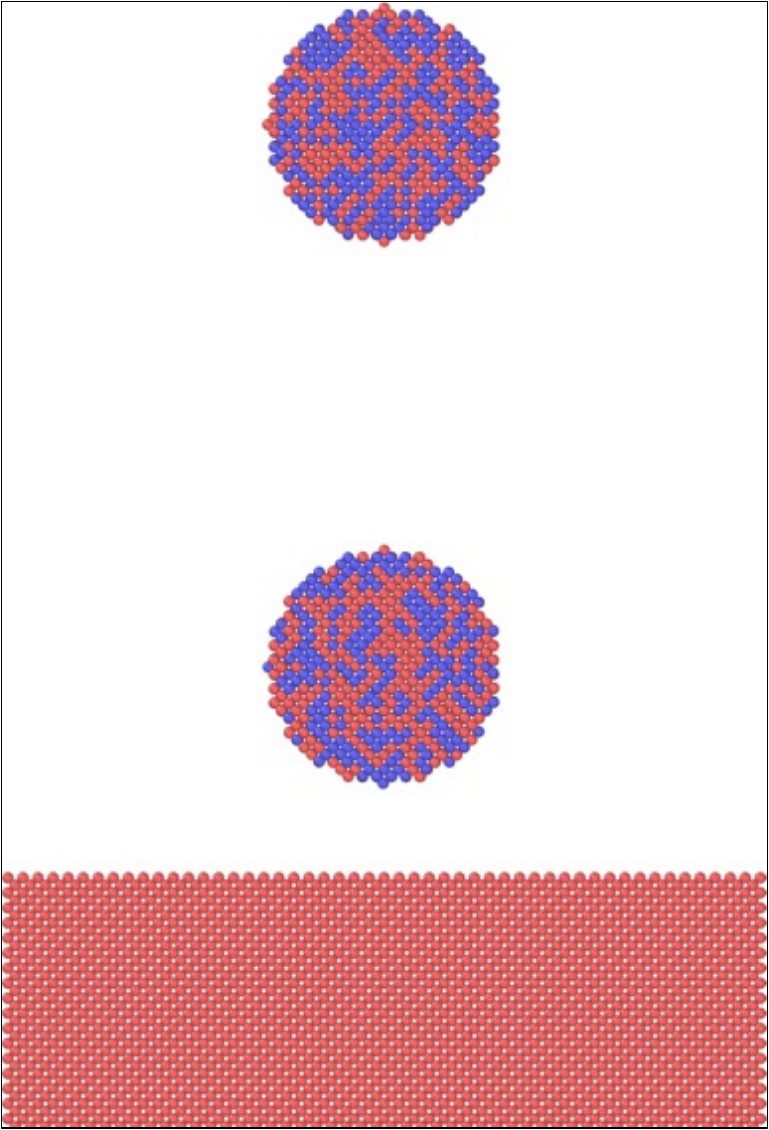}
    \label{fig:bef_v_w_anneal}
    }
    \subfloat[]
    {
\includegraphics[width=0.5\columnwidth]{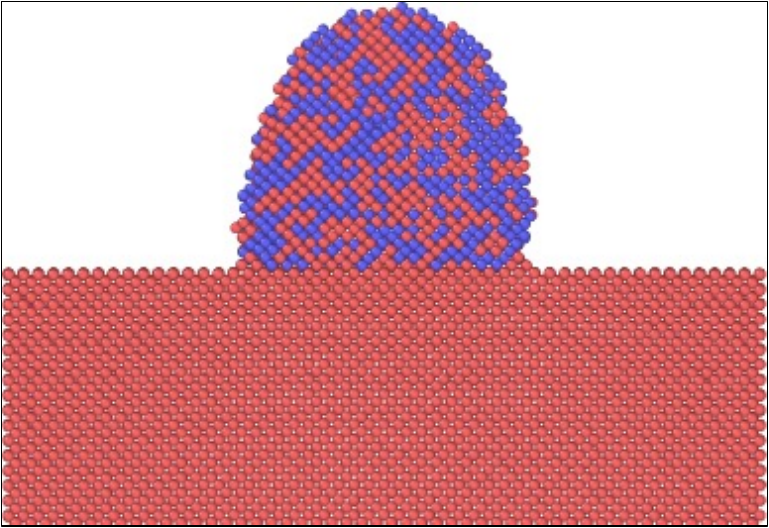}
        \label{fig:v_w_af}
    }
    \caption{Variation of atomic configuration \protect\subref{fig:bef_v_w_anneal} before impact at $t=0.1$ ns and, \protect\subref{fig:v_w_af} after subsequent impacts at $t=0.5$ ns with impact velocity of $1200$ \ms for equimolar $V$-$W$ particles where $V$ and $W$ are illustrated by blue and red atoms, respectively. 
    }
    \label{fig:alloy_config}
\end{figure}

\begin{figure}[t!]
    \centering
    \subfloat[]
    {
\includegraphics[width=0.5\textwidth]{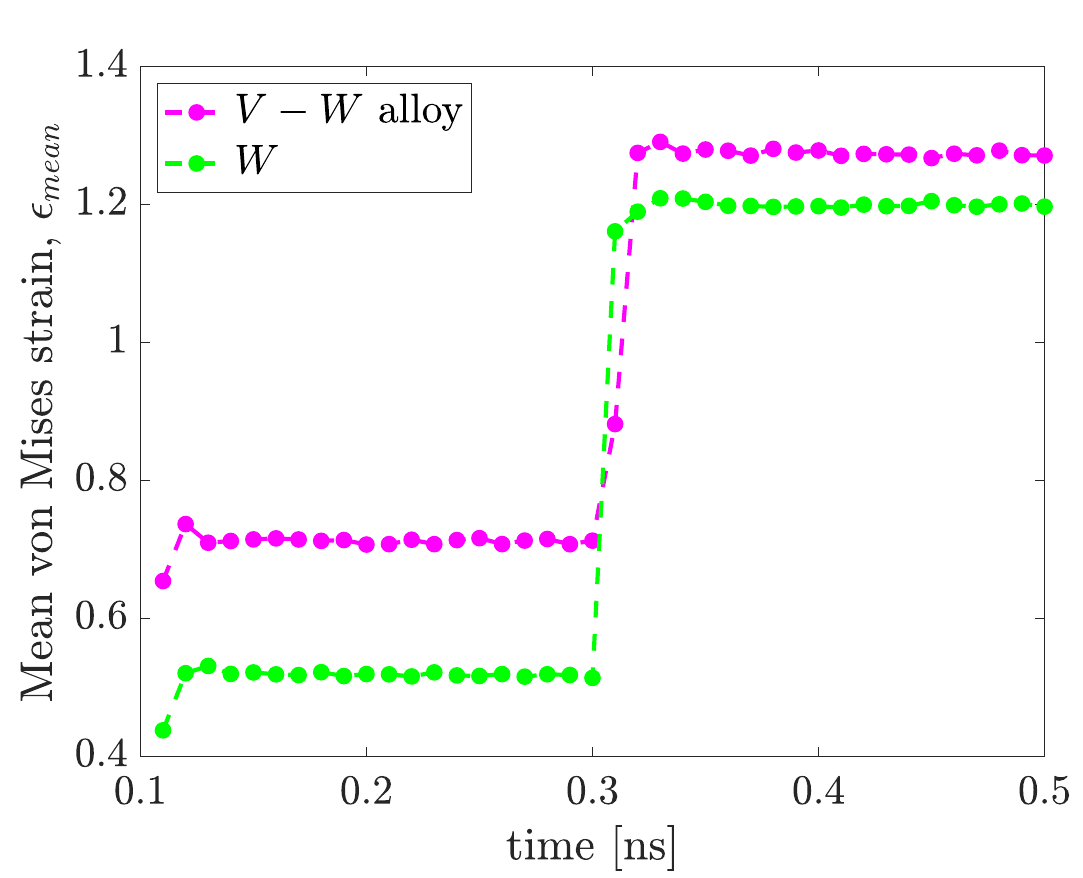}
    \label{fig:alloy_def}
    }
    \subfloat[]
    {
\includegraphics[width=0.5\columnwidth]{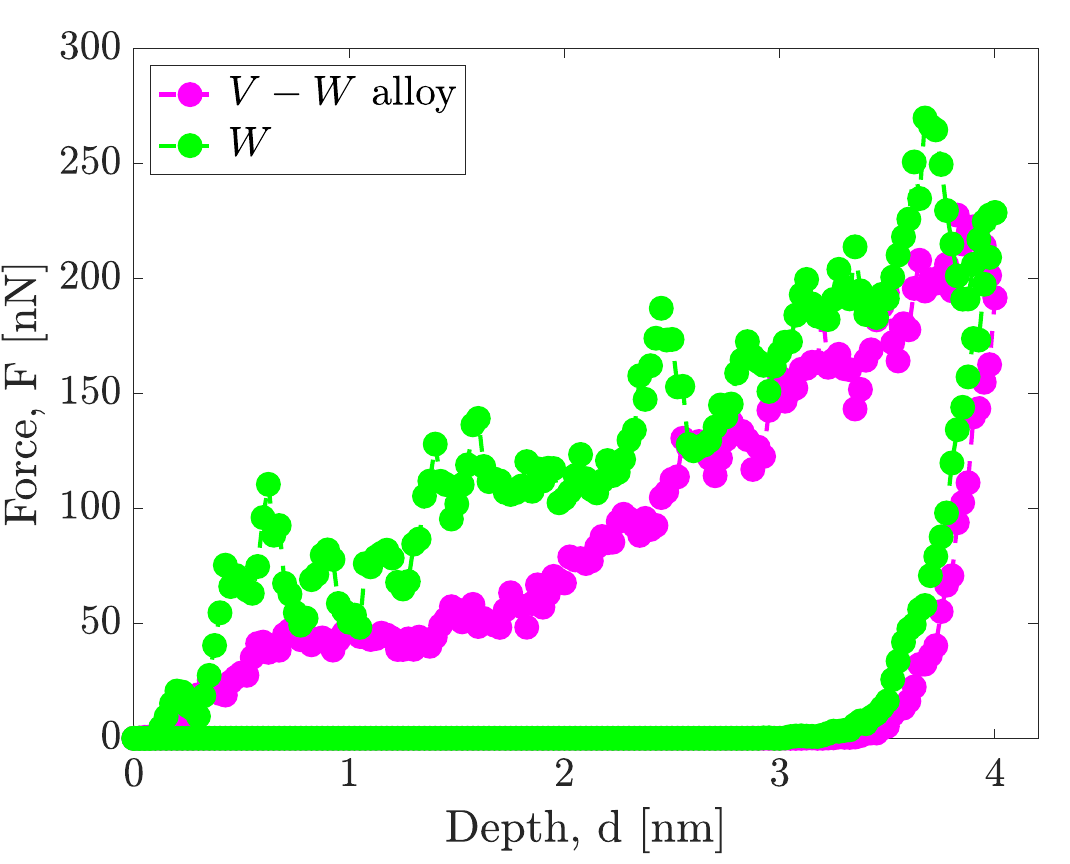}
        \label{fig:v_w_hard}
    } 
    \caption{Variation of \protect\subref{fig:alloy_def} mean von Mises strain as a function of time and \protect\subref{fig:v_w_hard} force as a function of nanoindentation depth for $W$ and equimolar $V$-$W$ alloy coating overlayer above $W$ substrate.}
    \label{fig:alloy_def_msd}
\end{figure}

To illustrate the deformation behavior, we considered two equimolar $V$-$W$ particles, with $V$ randomly dispersed in the $W$ matrix. We aim to utilize the considerable plastic deformation of particles and the substrate to form a $V$-$W$ alloy coating. Thus, a high impact velocity of $1200$ \ms is considered in this study. The initial configuration is shown in \fref{fig:bef_v_w_anneal}, where the bottom particle is subjected to $1200$ \ms at $t=0.1$ns, and the top particle is subjected to the same velocity at $t=0.3$ ns, similar to the configuration shown in \sref{sec:model_des}. As observed in \fref{fig:v_w_af}, there is an intermixing in between prealloyed $V$-$W$ particles, which causes the total vertical deformation of $45\si{\angstrom}$ for the particles due to CS impact. To compare the deformation of $V$-$W$ particles with that of $W$ particles during the CS process, we present the evolution of \emean as a function of time for both particles. \fref{fig:alloy_def} shows that the maximum value of \emean for the $V$-$W$ particle is approximately $9\%$ larger than the maximum value of \emean for the $W$ particle. This shows that the $V$-$W$ particle undergoes greater plastic deformation than the $W$ particle. The enhancement of plastic deformation in equimolar $V$-$W$ particle is a consequence of the softer nature of $V$-$W$ compared to pure $W$. The equimolar $V$-$W$ alloy coating is approximately $5\%$ shorter than the pure $W$ coating. This decrease in height corresponds to an increase in the coating's lateral spread. It is worth noting that a uniform lattice structure is absent throughout the coating. This non-uniformity can result from elastic stresses arising from a mismatch in bond lengths or from uncontrolled clustering of alloy components, leading to poor crystallinity and high spatial heterogeneity, including crystallographic distortions \citep{baranovskii2022energy}.

In addition to characterizing the coating layer's lateral spread, we aim to predict its mechanical properties. Below, we measure the hardness of both the $V$-$W$ alloy and the pure $W$ coating using atomic-scale nanoindentation simulations.

Nanoindentation simulations can provide quantitative information on the coating's hardness and bonding strength. Due to the indenter's penetration into the coating, local atomic rearrangements occur beneath the indenter tip that lead to the nucleation of dislocations, which mediates plastic deformation \cite{dominguez2024atomistic, mes2024atomistic}. The coating of a hard material always exhibits higher resistance to indentation, which resists the nucleation of dislocations pop-out \cite{ohmura2021pop}. To investigate whether the heterogeneous deposited layer of the $V$-$W$ alloy discussed above exhibits mechanical properties different from homogeneous $W$ coating on substrate material, $W$, we conducted nanoindentation simulations of the $W$ and $V$-$W$-coated $W$ substrate. The nanoindentation simulation samples have been prepared by cutting a few nanometers of coating from the top, leaving $2$ nm of flat coating above the substrate. We used a $50.64\si{\angstrom}$ diameter spherical indenter with force constant, $K=10$ eV\si{\angstrom}$^{-3}$. Nanoindentation was initiated at $X_3=180\si{\angstrom}$ and advanced to a total penetration depth of $4$ nm measured from the top surface in the $-X_3$ direction. The force-depth diagrams for $W$ and equimolar $V$-$W$ coating are shown in \fref{fig:v_w_hard}. The $V$-$W$ coating shows an immediate increase in force as the indenter is pushed downwards, whereas in pure $W$, the force starts increasing from zero after a time delay. Moreover, the slope for $W$ coating is approximately $20\%$ higher within the first $2$ nm depth than the equimolar $V$-$W$ alloy, thus reconfirming the reduction in hardness of the equimolar $V$-$W$ alloy compared to pure $W$, as also reported by \citet{dominguez2024plastic}.

\begin{figure}[t!]
    \centering
    \subfloat[]
    {
\includegraphics[width=0.5\textwidth]{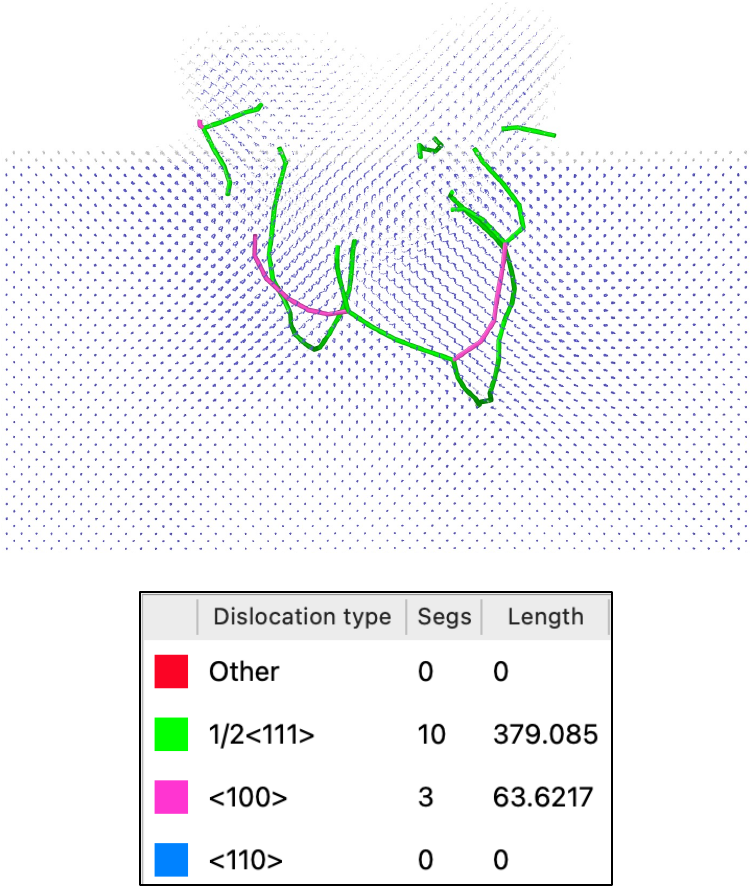}
    \label{fig:w_disloc}
    }
    \subfloat[]
    {
\includegraphics[width=0.5\columnwidth]{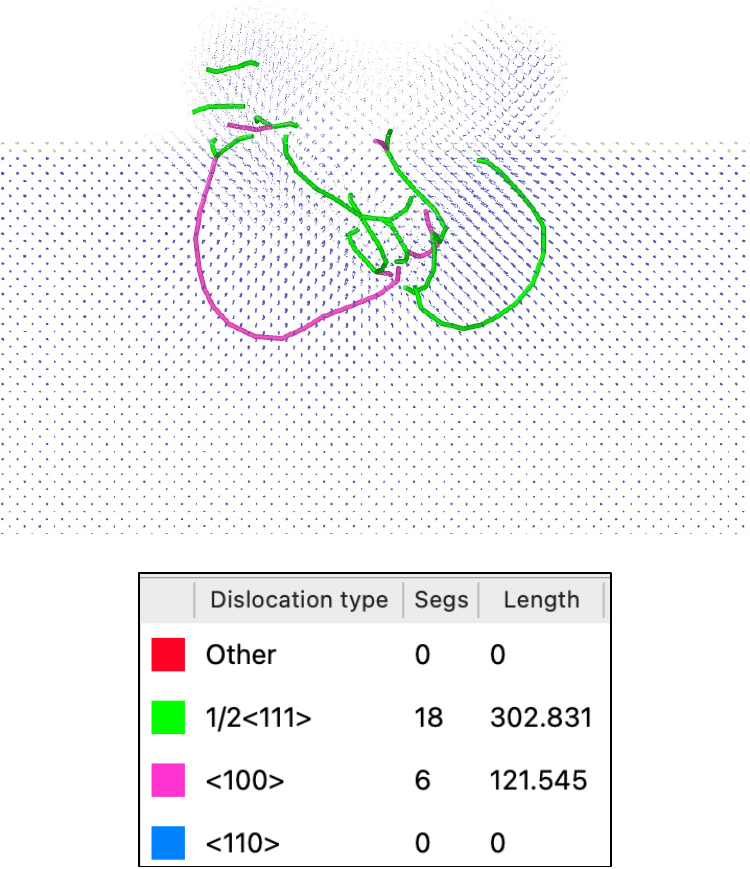}
        \label{fig:v_w_disloc}
    } 
    \caption{Dislocation lines in \protect\subref{fig:w_disloc} uncoated $W$ and \protect\subref{fig:v_w_disloc} $20\si{\angstrom}$ coated equimolar $V-W$ alloy over $W$ as the nanoindenter is $4$ nm beneath the top surface.}
    \label{fig:disloc_coat}
\end{figure}

One additional feature of nanoindentation of softer materials is the dislocation pop-up under load. To compare the dislocation densities for both $W$ and $V$-$W$ alloy-coated $W$ configurations, we plotted the dislocation analysis (DXA) results at a depth of $4$ nm from the top surface under load, computed using OVITO \citep{ovito}. As shown in \fref{fig:w_disloc}, the length of the total dislocation line is $\approx 442\si{\angstrom}$, while the length is $\approx 424\si{\angstrom}$ for the equimolar $V$-$W$ alloy as shown in \fref{fig:v_w_disloc}. While the DXA results show a relatively shorter total dislocation length for the $V$-$W$ coating, the total number of segments is almost twice that of the pure $W$ coating. This increment in the number of dislocation segments is an indicator of dislocation pop-up in relatively softer $V$-$W$ coating compared to $W$ \cite{ohmura2021pop,dominguez2024atomistic}. The changes in dislocation density, force-depth relation, and strain evolution for $V$-$W$ alloy are a clear indicator of the formation of a stable heterogeneous interface that has distinct behavior compared to a homogeneous interface. Thus, ultrasound-assisted CSAM provides a pathway to efficiently form heterogeneous interfaces through enhanced plastic deformation, eliminating the need for an external heating source and making it suitable for various on-site repair applications. 

\section{Summary and Conclusions}
\label{sec:summary}
In this paper, we investigated the role of ultrasound-assisted acoustoplasticity in cold spray (CS) particles using atomic-scale simulations. We demonstrated that ultrasonic perturbations induce pronounced acoustoplasticity through softening, enabling deformation mechanisms in tungsten that are otherwise inaccessible under conventional CS conditions. Acoustic softening facilitates a stress-assisted crystalline-to-amorphous transition accompanied by a transient temperature rise at the impact interface, leading to substantially enhanced plastic deformation during particle impact. This acoustoplasticity-driven response provides a viable pathway to achieve effective bonding and tailor microstructural evolution in refractory materials under extreme processing conditions. The key observations of this study are summarized below.

\begin{enumerate}
    \item Ultrasonic perturbations enhance maximum plastic deformation by approximately $1.5$ times across CS impact velocities ranging from $300$ \ms to $1200$ \ms, a behavior consistent with strong acoustic softening and reduced resistance to plastic flow.

    \item Ultrasonic perturbation refines grain microstructures as well as causes recrystallization due to the presence of elevated transient temperatures at the impact region, leading to the formation of fewer pores and defects compared to non-ultrasound-assisted configurations. The study shows grain boundary formation is observed mostly for particle diameters $>50.64\si{\angstrom}$.
    
    \item While an increase in ultrasonic perturbation amplitude increases the plastic deformation and associated flattening ratio, an increase in ultrasound frequency affects oppositely, decreasing both plastic deformation and flattening ratio due to the limited time available for the material to respond to constant agitation.

    \item Ultrasonic perturbation is utilized to form heterogeneous coating on $W$ substrate that can exhibit distinct properties and dislocation densities compared to the homogeneous coating on the base substrate material, depending on the particle composition.  
\end{enumerate}

Although cold spray additive manufacturing (CSAM) is widely used for on-site repair applications, its practical deployment remains largely limited to relatively soft materials such as nickel (Ni) and copper (Cu). This limitation significantly constrains the broader adoption of CSAM for extreme-environment applications. The present work demonstrates a viable pathway to extend the benefits of CSAM to hard and brittle refractory metals and to enable the formation of engineered alloys through ultrasound-assisted processing. In particular, ultrasonic excitation proves effective in promoting interfacial bonding and achieving uniform coatings, especially for larger CS particles where conventional CS struggles to induce sufficient plastic deformation.

Before concluding, it is important to acknowledge the limitations of this study. First, while atomistic simulations capture nanoscale deformation mechanisms, microstructural evolution, and transient thermal and mechanical responses during CS, the process inherently spans micro- to millimeter-length scales \citep{tianyu2022experimental,yang2024effect}. Consequently, quantities such as virial stress obtained from nanoscale simulations are not directly comparable to macroscopic stress measurements from experiments. Bridging this gap will require multiscale modeling frameworks that couple atomistic insights with continuum-scale descriptions. Second, collective effects such as multi-particle tamping during successive impacts were not considered. Finally, the present model does not include oxide layer formation at particle–substrate or particle–particle interfaces, which can strongly influence bonding and deformation in CS. Previous studies have shown that ultrasonic excitation can assist in disrupting and removing oxide layers \citep{long2019impacts}, suggesting an additional mechanism by which ultrasound may enhance bonding. Incorporating oxide dynamics and multi-particle interactions will be the focus of future work.

\section*{Acknowledgement}
Research was sponsored by the Office of Naval Research and was accomplished under Grant Number W911NF-25-1-0106. The authors would also like to acknowledge support from the U.S. Army Contracting Command – Aberdeen Proving Ground – Research Triangle Park Division, Award W911NF-25-2-0030. This research used the Delta advanced computing and data resource, which is supported by the National Science Foundation (award OAC 2005572) and the State of Illinois. The views and conclusions contained in this document are those of the authors and should not be interpreted as representing the official policies, either expressed or implied, of the Army Research Office or the U.S. Government. The U.S. Government is authorized to reproduce and distribute reprints for Government purposes, notwithstanding any copyright notation herein. 

\section*{CRediT author statement}
\noindent
\textbf{Md Tusher Ahmed:} Methodology, Software, Validation, Formal analysis, Investigation, Data Curation, Writing-Original Draft, Visualization.
\textbf{Farid Ahmed:} Conceptualization, Writing-Review \& Editing, Supervision, Funding acquisition.
\textbf{Jianzhi Li:} Conceptualization, Resources, Writing-Review \& Editing, Supervision, Funding acquisition.  

\bibliography{apssamp}

\end{document}